\documentclass[a4paper, 12pt]{article}
\usepackage[utf8x]{inputenc}
\usepackage{amsfonts}
\usepackage{amsmath}
\usepackage{amssymb}
\usepackage{amsthm}
\usepackage{booktabs}
\usepackage{bbm}
\usepackage{float}
\usepackage{multirow}
\usepackage{graphicx} 
\usepackage{graphics}
\usepackage[comma,authoryear,sort]{natbib}
\usepackage{longtable}
\usepackage[]{threeparttable}
\usepackage{rotating}
\usepackage{xcolor}
\usepackage{comment}
\usepackage{tikz}
\usepackage{eurosym}
\usepackage{textcomp}
\usepackage{makecell}
\usepackage{xurl}
\usepackage{float}

\usepackage{algorithm}	
\usepackage{algorithmic}

\usepackage{bibunits}

\defaultbibliography{arXiv}
\defaultbibliographystyle{chicago}

\usepackage[hidelinks, 
            implicit=false, 
            colorlinks = false,
            linkcolor = black,
            urlcolor  = black,
            citecolor = black,
            anchorcolor = black]{hyperref}

\def\tr{{\rm T}}

\DeclareMathOperator{\rvech}{rvech}

\def\diag{{\rm diag}}
\def\tr{{\rm tr}}

\def\bsS{{\boldsymbol \Sigma}}
\def\bsEta{{\boldsymbol \eta}}
\def\bsMu{{\boldsymbol \mu}}
\def\bsLambda{{\boldsymbol \lambda}}
\def\bsAlpha{{\boldsymbol \alpha}}
\def\bsBeta{{\boldsymbol \beta}}

\def\bsx{{\boldsymbol x}}

\newcommand{\overbar}[1]{\mkern 1.5mu\overline{\mkern-1.5mu#1\mkern-1.5mu}\mkern 1.5mu}

\title{Additive Covariance Matrix Models: Modelling Regional Electricity Net-Demand in Great Britain}
\author{
V. Gioia$\,^1$, M. Fasiolo$\,^2$, J. Browell$\,^3$, R. Bellio$\,^4$\\
$\,^1\,$\small University of Trieste, Department of Economics, Business, Mathematics and Statistics \\
$\,^2\,$\small University of Bristol, School of Mathematics\\
$\,^3\,$\small University of Glasgow, School of Mathematics and Statistics\\
$\,^4\,$\small University of Udine, Department of Economics and Statistics \\
\small vincenzo.gioia@units.it
}
\date{}

\def\spacingset#1{\renewcommand{\baselinestretch}
{#1}\small\normalsize} \spacingset{1}

\begin{document}

\begin{bibunit}

\maketitle

\begin{abstract}
\noindent
Forecasts of regional electricity net-demand, consumption minus embedded generation, are an essential input for 
reliable and economic power system operation, and energy trading.
While such forecasts are typically performed region by region, operations such as managing power flows require spatially coherent joint forecasts, which account for cross-regional dependencies. Here, we forecast the joint distribution of net-demand across the 14 regions constituting Great Britain's electricity network. Joint modelling is complicated by the fact that the net-demand variability within each region, and the dependencies between regions, vary with temporal, socio-economic and weather-related factors. We accommodate for these characteristics by proposing a multivariate Gaussian model based on a modified Cholesky parametrisation, which allows us to model each unconstrained parameter via an additive model. Given that the number of model parameters and covariates is large, we adopt a semi-automated approach to model selection, based on gradient boosting. In addition to comparing the forecasting performance of several versions of the proposed model with that of two non-Gaussian copula-based models, we visually explore the model output to interpret how the covariates affect net-demand variability and dependencies.

The code for reproducing the results in this paper is available at  
\url{https://doi.org/10.5281/zenodo.7315105}.
%
\noindent
\end{abstract}



\newpage

\section{Introduction}
\label{chap2:intro}

Electricity networks are changing from centralised systems, where power is generated by large power plants connected to the transmission network and consumed mostly on the distribution network, to decentralised networks where significant generation and storage is connected directly to distribution networks. The growth of this \textit{embedded generation} means that the transmission network now needs to serve the net-demand of customers, that is their demand net of local production. In Great Britain (GB), embedded production comes mostly from domestic and small-to-medium commercial solar and wind farms, as well as small thermal power plants. The lack of visibility of these units at the transmission level, combined with the weather-dependent nature of renewable generation leads to considerable challenges in energy trading and power system operation \citep{huxley2022uncertainties}. 

The purpose of this work is to support such key operations by proposing an \emph{interpretable} modelling approach that provides probabilistic, \emph{spatially coherent} short-term net-demand forecasts.
%
%
The energy industry is conservative by nature due to the need to maintain security of supply (\citealp[see Chapter 8 of ][]{von2006electric}). As a result, new processes will only be adopted if they are trusted, and interpretability plays an important role in building trust.
%
Further, interpretability is of critical importance when extrapolation is required, for example during exceptional events such as extreme temperatures, or when the model's predictions must be decomposed into the contribution of several effects. 

Predicting power flows on the electricity transmission network is a key motivating application for probabilistic, spatially coherent modelling in energy forecasting. This is important for both network operators, who are responsible for system security, and traders who must be aware of spatial variation in prices. Power flows are influenced by the injection and offtake of power from the network, as well as network configuration. They are also constrained by the physics of the network and must be forecasted to identify and mitigate any risk of exceeding thermal or stability limits. Therefore, spatial probabilistic forecasts of supply and demand are required to forecast power flows, and quantify uncertainty and risk associated with these constraints. Further, as the configuration of the network may change, any forecasting system must be flexible enough to allow the aggregation of supply and demand on the fly to calculate flows across relevant boundaries \citep{ tuinema2020probabilistic}. 

\begin{figure}[t]
\centering
\includegraphics[scale=0.5]{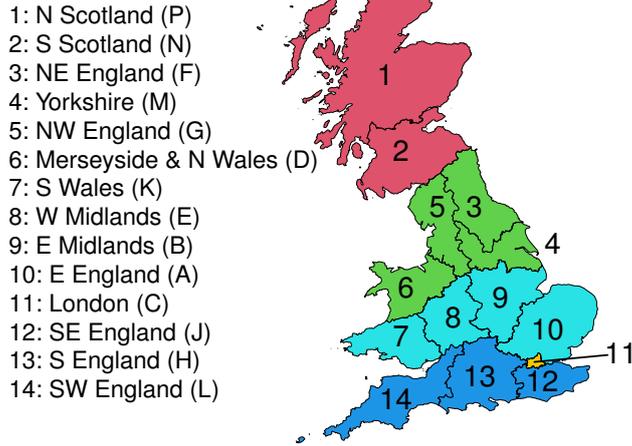}
\caption{A map of the GSP groups forming GB's electricity grid. The letters are the designation used by the electricity market in GB, while the numbers correspond to the position of each GSP group in the response vector, $\bold{y}_{i}$ (see Section \ref{sec:model_selection}). The colours represent five macro-regions namely Scotland (red), Northern (green), Midlands (light blue), Southern (blue) and London (yellow). The data on GSPs boundaries have been obtained from \protect\url{https://data.nationalgrideso.com}.}
\label{fig:GSP_groups}
\end{figure}

Motivated by the need for probabilistic joint net-demand forecasts, we consider joint modelling of net-demand across the 14 regions constituting GB's transmission network, which are shown in Figure \ref{fig:GSP_groups}. The net-demand in each region is the aggregate of the net-demand across many Grid Supply Points (GSPs), the latter being the interfaces between the transmission system and either a distribution network or a high-voltage consumer. 
Correctly modelling the dependency structure between regions is critical as an error in such a structure would pollute downstream predictions of power flows. Further, joint probabilistic forecasts of regional net-demand can be flexibly post-processed to produce forecasts tailored to the needs of different analyses. For example, when sustained winds blow in the North of the country, the boundary between Scotland and the North of England is of particular interest.  Indeed, the power flows across this boundary can be substantial, their direction and size being driven mostly by wind generation in Scotland, where embedded wind capacity far exceeds regional demand. Similarly, when the sun shines across the country, power flows from the South of England, where most of the country's solar generation units are installed, to London and the Midlands. During winter peak hours, the boundaries between London and its surroundings are characterised by heavy power flows directed toward the Capital, driven by high demand and low generation capacity in the city. 

While National Grid's 2021 ten-year statement \citep{elect10year} gives a detailed description of tens of transmission grid boundaries and explains under which circumstances the power lines crossing them become heavily loaded, the examples given above are meant to convey the fact that power flow analysis on operational time scales needs regional net-demand forecasts that can be aggregated, or more generally post-processed, to match the particular scenario of interest. For concreteness, in this work we consider the aggregation into the five GSP macro-regions shown in Figure \ref{fig:GSP_groups}, which are motivated by the boundaries mentioned above and match closely the critical boundaries presented in a Regional Trends and Insights report from the National Grid  (\citealp{NG_bound}, see Figure 1 therein).

\begin{figure}[t!]
\centering
\includegraphics[scale=0.22]{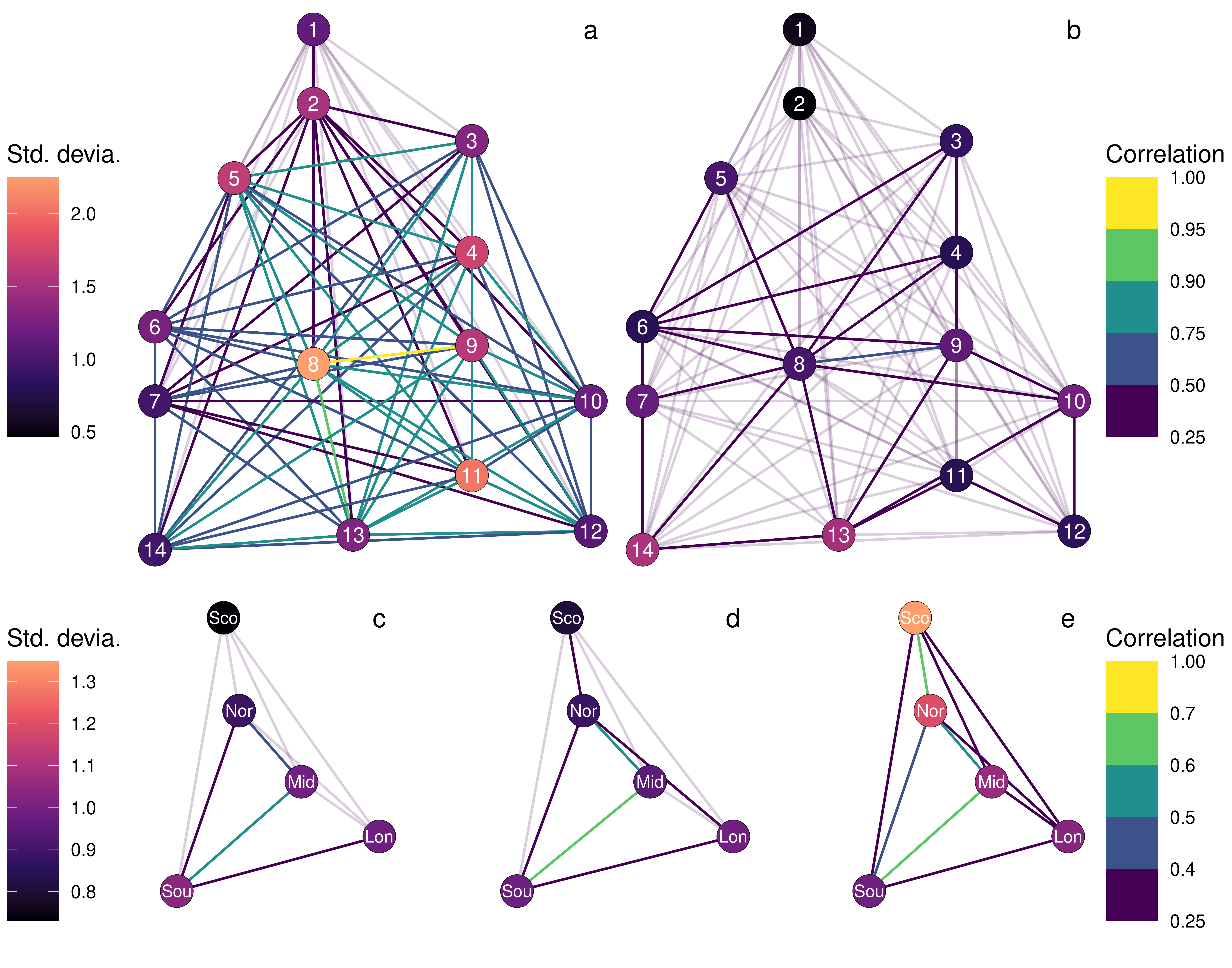}
\caption{Conditional standard deviations (nodes) and correlations (edges) across the 14 GSP regions (a--b) or macro-regions (c to e), predicted by the model from Section \ref{sec:select_results}. The edges corresponding to correlations lower than 0.25 have been made transparent. The plots correspond to 7am on 31/12/18 (a), midnight on 20/08/18 (b) and 10am on 14/06/18 (c to e). Plot d is based on regional wind and precipitation forecasts, while c and e correspond to, respectively, a 25\% decrease and increase of such forecasts. Note that the colour scales used for the nodes and the edges are different between a-b and c-e. 
}
\label{fig:corr_graph}
\end{figure}

Forecasts of power flows, which are composite variables of demand and supply, among many other factors, can only be calculated if a multivariate predictive distribution of said quantities is available. Hence, joint forecasts of net-demand across the GSP regions shown in Figure \ref{fig:GSP_groups}, or across an appropriate, scenario-dependent aggregation of them, are essential to help the network operator to take early action when managing the risk of breaching constraints.
 However, the structure of spatial dependency in net-demand is complex, as it is influenced by both socio-economic and weather effects, and is time-varying. Figure \ref{fig:corr_graph}a-b illustrates the issue. In particular, 
 the conditional regional standard deviations and inter-regional correlations of net-demand, predicted one day ahead by one of the models proposed in this paper. Figure \ref{fig:corr_graph}a corresponds to New Year's Eve, a day where net-demand forecast uncertainty is particularly high and correlation is strong between densely populated areas, such as London and the West Midlands. In contrast, Figure \ref{fig:corr_graph}b corresponds to the 20th of August and shows that net-demand is predicted to lead to a quieter day, from a network management perspective, with weak spatial dependency in forecast uncertainty.


Figure \ref{fig:corr_graph}a-b makes clear that capturing the time-varying nature of regional net-demand dynamics is essential to produce operationally useful joint forecasts. However, several other factors affect the joint distribution of regional net-demand, in addition to daily and yearly seasonalities. For example, the right column of Figure \ref{fig:corr_graph} shows the joint configuration of the macro-regional net-demand variabilities and dependencies during storm Hector. In particular, Figure \ref{fig:corr_graph}d shows the prediction obtained by conditioning on the seasonal factors and weather forecasts corresponding to this time period. Figures \ref{fig:corr_graph}c--e have been obtained by respectively decreasing and increasing the regional wind speed and precipitation forecasts by 25\%. They show that weather has a strong effect on the joint distribution of net-demand.
Specifically, a strengthening of the storm is predicted to lead to higher variability in Scotland and to stronger correlations between the latter and the other macro-regions.

Having motivated the need for an interpretable covariance matrix modelling framework able to provide the spatially coherent net-demand forecasts required by power flow analysis, we now outline the modelling approach proposed here.
We jointly model GB regional net-demand using a multivariate Gaussian model, based on a covariance matrix parametrisation that allows us to model each of its unconstrained parameters via a separate additive model, containing both parametric and smooth spline-based effects. In particular, the covariance matrix of the regional net-demand vector is parametrised via the modified Cholesky decomposition (MCD) of \cite{pourahmadi1999}. 
The wiggliness of the smooth effects is controlled via smoothing penalties, whose strength is controlled via smoothing parameters.  Model fitting is performed via two nested iterations, the regression coefficients being estimated via maximum a posteriori (MAP) methods, while the smoothing parameters are selected by maximising a Laplace approximation to the marginal likelihood (LAML). 

The proposed model can be seen as a multi-parameter generalized additive model \citep[GAM,][]{hastie1987generalized} or as a generalized additive model for location, scale and shape \citep[GAMLSS, ][]{rigby2005generalized}. 
Additive models are popular modelling tools in electricity demand forecasting (see, e.g., \citealp{fan2011short}), in part because they strike a balance between predictive performance and interpretability, the importance of the latter in this context having been discussed above. While ensuring interpretability is challenging under the richly parametrised model considered here, the MCD parametrisation provides some degree of interpretability when the response vector has some, not necessarily unique, intrinsic ordering, as is the case for regional net-demand. We further enhance the interpretable exploration of the model by summarising its output via the accumulated local effects (ALEs) of \cite{apley2020}.

The proposed joint regional net-demand model 
has 119 distributional parameters, controlling the mean vector and the covariance matrix of a conditional multivariate Gaussian distribution. Each parameter can be modelled via parametric and smooth effects of several covariates, hence the space of possible models is large. While the effects controlling the mean vector can be chosen based on expert knowledge or previous research, manual selection of an additive model for each of the remaining 105 parameters is unrealistic. Here, we leverage the interpretation of the MCD's parameters to choose 
the set of candidate effects that could be used to model each parameter. Then, we use gradient boosting \citep{friedman2001greedy} to order the effects on the basis of how much they improve the fit, and we choose the number of effects modelling the MCD elements by maximising the forecasting performance on a validation set. The results show that the semi-automatic effect selection procedure just outlined leads to satisfactory predictive performance and to model selection decisions that are largely in agreement with intuition (e.g., wind speed and solar irradiance are selected to model net-demand variability in, respectively, Scotland and the South of England). 

To our knowledge, this is the first applied statistical paper to consider additive modelling of each parameter of the mean vector and of an unconstrained covariance matrix parametrisation, in a context where the response vectors are not low-dimensional 
and have heterogeneous elements, i.e. they are not lagged values of the same variable. Additive modelling of multivariate responses has been proposed by \cite{klein2015bayesian}, who consider bivariate Gaussian and t-distributions based on a variance-correlation decomposition. \cite{marra2017bivariate} propose a fitting framework where bivariate responses are modelled via copulas with continuous margins, and all distributional parameters are modelled additively. Also relevant are the covariate-dependent copula approaches of \cite{vatter2018} and \cite{hans2022boosting}, who provide examples featuring respectively four- and two-dimensional responses. Similarly to \cite{hans2022boosting}, \cite{stromer2023boosting} use the gradient boosting methods of \cite{thomas2018gradient} for model fitting, and consider several two-dimensional response models, including the bivariate Gaussian one.  \cite{klein2022multivariate} propose a copula-related approach that makes minimal distributional assumptions on the margins. While their model did not originally include penalised smooth effects, which are essential for the application considered here, it is conceivable that such effects could be included. Such an addition, if supported by sufficiently scalable fitting methods and software, could make their approach a serious alternative to our proposal. 

In a generalized linear modelling (GLM) context, \cite{pourahmadi1999} uses the MCD to parametrise a multivariate Gaussian model in eleven dimensions. They are interested in capturing temporal dependencies in longitudinal data, which allows them to impose a strong structure on the covariance matrix model. 
Beyond the bivariate case, \cite{bonat2016multivariate} model directly the covariance matrix using covariates and tune the optimiser to avoid generating indefinite matrices, while \cite{browell2022} use covariates-dependent covariance functions which, in some cases, do not provide positive definiteness guarantees and require perturbing the estimated covariance matrix to achieve it.

From a methodological perspective, our work is closely related to \cite{muschinski2024}, who consider non-parametric regression with multivariate Gaussian responses. They propose modelling of the elements of several Cholesky-based parametrisations, including the MCD. But, they fit the model using MCMC methods, rather than direct optimisation methods as done here and, while they compare a set of manually-chosen covariance matrix models on a weather forecasting application, here we consider semi-automatic variable selection to handle a much larger set of candidate covariates. Further, they aim at capturing temporal rather than spatial dependencies, the latter being the focus of this work.

The rest of the paper is structured as follows. Section \ref{chap2:model} introduces, in a general setting, the proposed multivariate Gaussian model structure and fitting methodology. It also summarises the inferential framework and motivates the use of ALEs for model output exploration. Section \ref{sec:application} focuses on the regional net-demand modelling application. In particular, the data is introduced in Section \ref{ref:data_description}, while Section \ref{sec:model_selection} describes the bespoke, boosting-based model selection approach proposed here. The output of the final model is explored in Section \ref{sec:select_results}, while the forecasting performance of the proposed model is assessed in Sections \ref{sec:GSP14_eval} and  \ref{sec:GSP14_eval_group}.  Section \ref{sec:conclusions} summarises the main results.

\section{Multivariate Gaussian Additive Models}
\label{chap2:model}

\subsection{Model Structure} \label{sec:mod_struct}
Let $\bold{y}_i=(y_{i1}, \ldots, y_{id})^\top$, for $i=1, \ldots,n$, be independent response vectors, normally distributed with mean $\boldsymbol \mu_i$ and covariance matrix $\bsS_i$. The $q = d + d(d+1)/2$ unique elements of $\boldsymbol \mu_i$ and of (a suitable parametrisation of) $\boldsymbol \Sigma_i$
are modelled via $\boldsymbol \eta_i$, a $q$-dimensional vector of linear predictors. The $j$-th element of $\boldsymbol \eta_i$ is modelled via
\begin{equation} \label{eq:basicGAM}
    \eta_{ij}={\mathbf{Z}^j_{i}}^\top \boldsymbol{\psi}_j + \sum_{l} f_{jl}({\boldsymbol x}^{S_{jl}}_i)\,\,, \;\;\; \text{for} \;\;\; j = 1, \dots, q,
\end{equation}
where ${\mathbf{Z}^j_{i}}^\top$ is the $i$-th row of the design matrix $\mathbf{Z}^j$, $\boldsymbol \psi_j$ is a vector of regression coefficients, $\boldsymbol x_i$ is an $s$-dimensional vector of covariates and $S_{jl} \subset \{1, \dots, s\}$. Hence, for example, if $S_{jl} = \{2, 4\}$ then ${\boldsymbol x}_{i}^{S_{jl}}$ is a two dimensional vector formed by the second and fourth element of $\boldsymbol x_i$. Each $f_{jl}$ is a smooth function, built via
\begin{equation}
    f_{jl}(\boldsymbol x^{S_{jl}}) = \sum_{k} b^{jl}_k (\boldsymbol x^{S_{jl}}) \alpha_k^{jl}\,\,,
\end{equation}
where $b^{jl}_k$ are spline basis functions of dimension $\text{card}(S_{jl})$, while $\alpha_k^{jl}$ are regression coefficients. Denote with $\boldsymbol \alpha$ the vector of all such coefficients in the model. The wiggliness of the effects is controlled by an improper multivariate Gaussian prior on $\boldsymbol \alpha$. The prior is centered at the origin and its precision matrix is ${\bf S}^{\boldsymbol \lambda} = \sum_{u} \lambda_u {\bf S}_u$, where the ${\bf S}_u$'s are positive semi-definite matrices and $\boldsymbol \lambda = (\lambda_1, \lambda_2, \dots)^\top$ is a vector of positive smoothing parameters. See \cite{Wood2017} for a detailed introduction to GAMs, smoothing splines bases and penalties.

Let us temporarily drop index $i$ to simplify the notation. In this work we use the following parametrisation of $\boldsymbol \mu$ and $\boldsymbol \Sigma$ in terms of $\boldsymbol \eta$: $\mu_j = \eta_j$ for $j = 1, \dots, d$, while the remaining elements of $\boldsymbol \eta$ parametrise an MCD of $\boldsymbol \Sigma^{-1}$ \citep{pourahmadi1999}. In particular, 
\begin{equation}\label{mcd_prec}
\boldsymbol{\Sigma}^{-1}=\bold{T}^\top\bold{D}^{-2}\bold{T} \, ,
\end{equation}
where $\bold{D}^2$ is a diagonal matrix with 
${\rm D}^2_{jj}=\exp(\eta_{j+d})$, for $j = 1, \dots, d$,  and 
\begin{equation}\label{eq:mcd_T}
\bold T=\begin{pmatrix} 1 & 0 & 0 & \cdots& 0 \\
                               \eta_{2d+1} & 1 &  0 & \cdots& 0  \\

                               \eta_{2d+2} & \eta_{2d+3} & 1& \cdots   & 0 \\

                               \vdots & \vdots & \vdots & \ddots   &  \vdots&  \\

                             \eta_{q-d+2} & \eta_{q-d+3} & \cdots &  \eta_{q} & 1\\

\end{pmatrix}.
\end{equation}

Parametrisation (\ref{mcd_prec}) is unconstrained, that is the resulting covariance matrix $\bsS$ is positive definite for any finite $\bsEta$, which facilitates model fitting. Other unconstrained parametrisations could have been used, such as those discussed by \cite{pinheiro1996} and \cite{pourahmadi2011}, but the MCD approach is particularly attractive in the context of this work. First, the fitting methods described in Section \ref{sec:model_fitting} require the first two derivatives of the log-likelihood w.r.t. $\bsEta$ and, under the MCD parametrisation, the multivariate Gaussian log-likelihood can be written directly in terms of $\bsEta$, which eases the computation of such derivatives. 
Second, the MCD parametrisation has a regression-related interpretation, which can be exploited when the response vector has some intrinsic ordering.  In particular, assume w.l.o.g. that $\mathbb{E}(\bold y) = \bold 0$ and that $\bold y$ follows the regression models
$$y_{l} = \sum_{k=1}^{l-1}\phi_{lk}y_{k} + \epsilon_{l}\,\,, \quad \text{for} \quad l = 2, \dots, d,$$
where $y_1 = \epsilon_1$, $\text{var}(\epsilon_{l}) = \sigma_{l}^2$ and $\text{cov}(\epsilon_l, \epsilon_k) = 0$ for $l\neq k$. \cite{pourahmadi1999} shows that ${\rm T}_{lk} = -\phi_{lk}$  and ${\rm D}_{kk}^2 = \sigma_k^2$, for $k=1,\dots, d$, and $l = k+1, \dots, d$. Hence, the elements of $\bold T$ can be interpreted as the regression coefficients of the elements of $\bold y$ on their predecessors, while the non-zero elements of $\bold{D}^2$ are the residual variances of such regressions. 

Note that the interpretation of the MCD elements depends on the ordering of the elements of the response vector. Hence, the MCD parametrisation is particularly attractive when the response vector has some natural ordering, as is the case when dealing with chronologically ordered responses. While in this work we consider a response vector that does not have a unique natural ordering, in Section \ref{sec:model_selection} we will discuss how the spatial nature of GB regional net-demand data allows us to exploit the interpretation of the MCD parametrisation to guide the development of a multivariate model.  

\subsection{Model Fitting}\label{sec:model_fitting}

Let us indicate the set of all response vectors $\bold y_1, \dots, \bold y_n$ simply with $\bold y$ and with $\bsBeta$ the vector of all regression coefficients in the model, which include $\bsAlpha$ and all the unpenalised coefficients vectors $\boldsymbol{\psi}_j$. Let $\tilde{\bf S}^{\boldsymbol \lambda}$ be the prior precision matrix of $\bsBeta$, that is an enlarged version of ${\bf S}^{\boldsymbol \lambda}$ padded with zeros so that $\bsAlpha^\top {\bf S}^{\bsLambda} \bsAlpha=\bsBeta^\top \tilde{\bf S}^{\boldsymbol \lambda} \bsBeta$. Then, up to an additive constant that does not depend on $\bsBeta$, the Bayesian posterior log-density of the model from Section \ref{sec:mod_struct} is
\begin{equation} \label{eq:log_posterior}
 \mathcal{L}(\bsBeta) = \log p(\bsBeta|\bold y, \boldsymbol \lambda) = \sum_{i=1}^n \log p(\bold y_i|\bsBeta) - \frac{1}{2} \bsBeta^\top \tilde{\bf S}^{\boldsymbol \lambda} \bsBeta,  
\end{equation}
where $\log p(\bold y_i|\bsBeta)$ is the $i$-th log-likelihood contribution. 

For fixed smoothing parameters, $\boldsymbol \lambda$, we obtain MAP estimates of the regression coefficients by maximising the log-posterior (\ref{eq:log_posterior}), using Newton's algorithm. The latter requires the gradient and Hessian of the log-posterior w.r.t. $\bsBeta$, which are provided in the Supplementary Material \ref{AppA1} (henceforth SM \ref{AppA1}). The real challenge is selecting the smoothing parameters themselves.
We do it by maximising an approximation to the log marginal likelihood, $\mathcal{V}(\bsLambda) =   \log \int p(\bold y| \bsBeta)p(\bsBeta|\bsLambda) d \bsBeta$. In particular, we consider the LAML criterion
\begin{equation} \label{eq:LAML}
\tilde{\mathcal{V}}(\bsLambda) =  \mathcal{L}(\hat{\bsBeta})+\frac{1}{2}\log|\tilde{\bf S}^{\bsLambda}|_{+}-\frac{1}{2}\log|\boldsymbol {\mathcal{H}}|+\frac{M_{p}}{2}\log(2\pi)\,\,,
\end{equation}
with $M_{p}$ being the dimension of the null space of $\tilde{\bf S}^{\lambda}$, $|\tilde{\bf S}^{\lambda}|_{+}$ the product of its positive eigenvalues,   $\hat{\bsBeta}$ the maximiser of $\mathcal{L}(\bsBeta)$ and $\boldsymbol{\mathcal{H}}$ its negative Hessian, evaluated at $\hat{\bsBeta}$.

We maximise $\tilde{\mathcal{V}}(\bsLambda)$ via the generalized Fellner-Schall method of \cite{woodfasiolo2017}, under which the $u$-th smoothing parameter is updated using
\begin{equation} \label{eq:FS}
\lambda^{\text{new}}_u = \frac{\tr\{(\tilde{\bold S}^{\boldsymbol \lambda})^{-} \tilde{\bold S}_u\}-\tr(\boldsymbol{\mathcal{H}}^{-1}\tilde{\bold S}_u)}{\hat {\bsBeta}^\top \tilde{\bold S}_u \hat {\bsBeta}}\lambda_u^{\text{old}} \,\, ,
\end{equation}
where $(\tilde{\bold S}^{\boldsymbol \lambda})^{-}$ is the Moore-Penrose pseudoinverse of $\tilde{\bold S}^{\boldsymbol \lambda}$ and $\tilde{\bold S}_u$ is ${\bold S}_u$ after padding it with zeros. That is, if we indicate with $\bsBeta_{u}$ the subvector of $\bsBeta$ that is penalised by ${\bold S}_u$, then $\bsBeta_{u}^\top{\bold S}_u\bsBeta_{u} = \bsBeta^\top\tilde{\bold S}_u\bsBeta$. An advantage of update (\ref{eq:FS}) is that it does not require computing the derivatives of $\tilde{\mathcal{V}}(\bsLambda)$ w.r.t. $\bsLambda$. In particular, as detailed in \cite{wood2016}, computing the gradient of $\tilde{\mathcal{V}}(\bsLambda)$ requires the third derivatives of log-likelihood w.r.t. each element of $\bsEta$, which leads to computational effort of order $O\left\{n\binom{q+2}{3}\right\}$ ($\approx 2\times 10^{10}$, if $q = 119$ and $n \approx 8\times10^4$ as in the application considered here). Hence, for moderately large dimension $d$ of the response vector, a quasi-Newton iteration for maximising $\tilde{\mathcal{V}}(\bsLambda)$ would be too computationally intensive, at least under na\"ive evaluation of the likelihood derivatives. 

\subsection{Inference and Effect Visualisation} \label{sec:inference}
The uncertainty of the fitted regression coefficients, $\bsBeta$, can be quantified via the approximate Bayesian methods detailed in \cite{wood2016}, which we summarise here. In particular, standard Bayesian asymptotics justify approximating $p(\bsBeta|\bold y, \bsLambda)$ with a Gaussian distribution, $N(\hat{\bsBeta}, {\bf V}_{\bsBeta})$, centered at the MAP estimate and with covariance matrix ${\bf V}_{\bsBeta} = -\boldsymbol {\mathcal{H}}^{-1}$. Such a posterior approximation does not take into account the uncertainty of the smoothing parameters estimates, which are considered fixed to the LAML maximiser. \cite{wood2016} use a Gaussian approximation to $p(\bsLambda|\bold y)$ and propagate forward the corresponding smoothing parameter uncertainty to obtain an approximation to the unconditional posterior, $p(\bsBeta|\bold y)$. In principle, this approach could be adopted for the model class considered here, but the formulae provided by \cite{wood2016} require the Hessian of $\tilde{\mathcal{V}}$ w.r.t. $\bsLambda$ which involves the fourth derivative of log-likelihood w.r.t. each element of $\bsEta$. While it might be possible to reduce the analytical effort needed to obtain such derivatives by automatic differentiation  \cite[see, e.g.,][]{griewank2008evaluating} the computational cost mentioned in Section \ref{sec:model_fitting} would still be an obstacle.

Given that the smooth effects are linear combinations of the regression coefficients, it is simple to derive pointwise Bayesian credible intervals for the effects, the asymptotic frequentist properties of such intervals having been studied by \cite{nychka1988bayesian}. However, each effect acts directly on a linear predictor, the latter being non-linearly related to one or more elements of $\bsS$. As explained in Section \ref{sec:mod_struct}, the MCD parametrisation is related to a set of regressions involving the elements of the response vector.  
This fact aids interpretability only if the response vector has some natural ordering. While this is to some extent the case in the application considered here (see Section \ref{sec:model_selection}), communicating modelling results to non-statisticians is more likely to be effective if framed in terms of widely-used concepts such as covariances and correlations, rather than parametrisation-specific quantities. Hence, we use the accumulated local effects (ALEs) of \cite{apley2020} to quantify the effect of a covariate on $\bsS$ or on the corresponding correlation matrix, $\boldsymbol \Gamma$. 

In contrast to the partial dependence plots \citep{friedman2001greedy}, ALEs avoid making an extrapolation error when the covariates are correlated. This is explained by \cite{apley2020}, who also provide formulas for estimating ALEs, and quantify their uncertainty via bootstrapping. Here, we exploit the results of \cite{capezza2021additive} for multi-parameter GAMs to approximate the posterior variance of ALEs. In SM \ref{app:jacobians} we provide more details on ALEs, while in Section \ref{sec:select_results} we use them to visualise a model for GB regional net-demand.

\section{Joint Multivariate Regional Net-Demand Modelling} \label{sec:application}

\subsection{Data Description and Modelling Setting} \label{ref:data_description}

We consider data on regional net-demand in GB, from five years, 2014 to 2018. Net-demand is the load measured at the interface between transmission and distribution networks. In GB these interfaces are called Grid Supply Points (GSPs) and are grouped into 14 GSP groups. Let $y_{ij}$, for $i = 1,\dots, n$, be the standardised net-demand of GSP group $j$ measured at a 30min resolution. In addition to net-demand, the data contain the covariates listed in Table \ref{tab:Covariates} or transformations thereof. Some covariates are common to all GSP groups, while others are region-specific, such as those derived from the hourly day-ahead weather forecasts produced by the operational ECMWF-HRES model. Gridded weather predictions are summarised via the regional features reported in Table \ref{tab:Covariates}.

\begin{table}[t]
\footnotesize
\renewcommand{\arraystretch}{1.29}
\begin{center}  
\begin{tabular}{ | l @{\hskip 0.08cm}  p{6.75cm}   | l @{\hskip 0.08cm} p{6.60cm}  |} 
\hline  
\multicolumn{2}{|l|}{General covariates} & \multicolumn{2}{l|}{Covariates derived from weather forecasts} \\  \hline 
 $\text{dow}_{i}$ & day of the week factor
& $\text{rain}_{ij}$ & mean precipitation (m/day) \\ 
$\text{dow}^+_{ij}$ & $\text{dow}_{i}$ with additional factor levels accounting for public holidays 
& $\text{temp}_{ij}$ & temperature (K) at cell with highest regional population density \\
$\text{t}_i$ & time since the 1st January 2014
& $\text{temp}_{ij}^S$ & 48 hours rolling mean of $\text{temp}_{ij}$ \\   
$\text{shol}_{ij}$ & school holidays, three levels factor to distinguish Christmas from other holidays 
& $\text{irr}_{ij}$ & mean solar irradiance (W/$\text{m}^2$) times embedded solar generation capacity (MW) \\
$\text{tod}_i$ & time of day ($ \in \{0, 0.5, \dots, 23.5\}$)
& $\text{wsp}^{10}_{ij}$ & mean wind speed at $10$ meters (m/s) \\ 
$\text{wcap}_i$ & GB embedded wind generation capacity (MW)
& $\text{wsp}^{100}_{ij}$ & mean wind speed at $100$ meters (m/s) \\ 
$\text{doy}_{i}$ & day of the year ($\in \{1, \dots, 366\}$)
&  & \\ 

$\text{n2ex}_i$ & N2EX day-ahead electricity price (\textsterling /MWh)
&  & \\ 
$y_{ij}^{24}$ & net-demand at a 24 hours lag 
&  & 
\\ 
\hline     
\end{tabular}  
\caption{Covariates used to model GB regional net-demand.}
\label{tab:Covariates}
\end{center} 
\end{table}

\cite{browell2021probabilistic} use the data just described to model the conditional distribution of $y_{ij}$, separately for each of the $d=14$ regions. They do so using a composite modelling approach, where the raw residuals of a Gaussian GAM are modelled via linear quantile regression. Extreme quantiles are modelled using a GAMLSS model based on the generalized Pareto distribution. Here, we are interested in modelling the joint distribution of the $d$-dimensional response vector $\bold y_{i}$. 
We consider a multivariate Gaussian model $\bold y_{i} \sim N(\bsMu_{i}, \bsS_{i})$, where $\bsMu_{i}$ and $\bsS_{i}$ are controlled by the linear predictors vector $\bsEta_{i}$, as described in Section \ref{sec:mod_struct}, and each element of $\bsEta_{i}$ is modelled via (\ref{eq:basicGAM}). 

The proposed model has $q = 119$ linear predictors and each of them could be modelled via any of the covariates described above. Hence, model selection is challenging. 
As explained in 
Section \ref{sec:model_selection}, we use 2014-2016 data to generate a long list of candidate covariate effects, ordered in terms of decreasing importance. We choose the number of effects to add to the final multivariate Gaussian model, that is where to stop along the ordered effect list, by maximising the out-of-sample predictive performance on 2017 net-demand. Having chosen the model structure, in Section \ref{sec:select_results} we explore the model output, and in Sections \ref{sec:GSP14_eval} and \ref{sec:GSP14_eval_group} we evaluate the accuracy of the resulting forecasts on 2018 data.

\subsection{Semi-Automatic Model Selection} \label{sec:model_selection}

\cite{browell2021probabilistic} consider a progression of univariate GAMs based on an increasingly rich set of covariates and assess their performance on a day-ahead forecasting task. We use their results to choose a model for the first $d = 14$ elements of $\bsEta_{i}$ or, equivalently, $\bsMu_i$. In particular, we adopt the model formula
\begin{align}\label{mean_formula}
\eta_{ij} &= g_{j1}(t_i) + g_{j2}(t_i^2) + g_{j3}(\text{dow}^+_{ij}) + g_{j4}(\text{shol}_{ij}) +  g_{j5}(y_{ij}^{24}) + g_{j6}(\text{wsp}_{ij}^{10}) \nonumber \\ & + f_{j1}^{20}(\text{doy}_i) + f_{j2}^{35}(\text{tod}_i) + f_{j3}^{10}(\text{n2ex}_i) + f_{j4}^{35}(\text{temp}_{ij}) + f_{j5}^{35}(\text{temp}_{ij}^S) + f_{j6}^{10}(\text{rain}_{ij}^{1/2}) \nonumber \\ & + \text{wcap}_i \times f_{j7}^{20}(\text{wsp}_{ij}^{100}) + f_{j8}^{5}(\text{irr}_{ij}) + f_{j9}^{30}(\text{tod}_{i}, \text{dow}^+_{ij}) + f_{j10}^{20}(\text{tod}_{i}, \text{shol}_{ij}) \nonumber \\ & + f_{j11}^{5, 5}(\text{n2ex}_i, \text{tod}_{i}) + f_{j12}^{5, 5}(\text{temp}_{ij}, \text{tod}_{i}) + f_{j13}^{5, 5}(\text{rain}_{ij}^{1/2}, \text{tod}_{i}) + f_{j14}^{10, 10}(\text{doy}_i, \text{tod}_{i})\,\,,
\end{align}
for $j = 1, \dots, d$. Here $g_{j1}$ to $g_{j6}$ are parametric (linear) effects, while $f_{j1}$ to $f_{j14}$ are smooth effects. In particular, $f_{j1}$ to $f_{j8}$ are univariate smooth effects, the spline bases dimensions being indicated by the superscripts. Effects $f_{j9}$ and $f_{j10}$ are smooth-factor interactions, where a different univariate smooth is defined for each level of the $\text{dow}^+_{ij}$ or $\text{shol}_{ij}$ factor variables. The last four effects in (\ref{mean_formula}) are bivariate tensor-product smooths, where the dimension of each marginal basis is indicated by the superscripts. All smooth effects are built using cubic regression spline bases, except for $f_{j1}^{20}(\text{doy}_i)$, which uses a B-spline basis with an adaptive P-spline penalty. The latter allows the smoothness of the effect to vary with $\text{doy}_i$, see Section 5.3.5 of \cite{Wood2017} for details. Model (\ref{mean_formula}) is similar to the ``GAM-point'' model of \cite{browell2021probabilistic} but their model lacks the interaction between $\text{doy}_i$ and $\text{tod}_{i}$, and uses $\text{rain}_{ij}$ rather than $\text{rain}_{ij}^{1/2}$, the transformed version leading to more even use of basis functions (rainfall is rightly skewed).
Further, they use a parametric effect, based on basic trigonometric functions of $\text{doy}_i$, to model the annual seasonality. The approach used here models seasonality more flexibly, which is particularly important around year-end and in densely populated regions, such as London (see the results in Section \ref{sec:select_results}). 

It is challenging to develop a model for the remaining 105 elements of $\bsEta_{i}$. As explained in Section \ref{sec:mod_struct}, the elements of the $\bold T_i$ and $\bold D_i$ matrices correspond to the parameters of a set of linear models, where the $j$-th element of $\bold y_i$ is regressed on its predecessors $y_{ij-1}$, \dots, $y_{i1}$. Hence, the parameters of the decomposition depend on the ordering of the elements of $\bold y_i$. For the GSP net-demand data, a sensible ordering can be chosen based on the location of the GSP regions. As Figure \ref{fig:GSP_groups} shows, we order the regions North to South hence $y_{i1}$ and $y_{i14}$ are, respectively, net-demand in the North of Scotland and in the South West of England. Under such an ordering, neighbouring regions, which are more likely to be affected by similar weather and socio-economic events, occupy nearby positions in $\bold y_i$. Of course, variations on the proposed ordering could be considered, for example one could think about swapping the order of regions 12 and 13, which are at a similar latitude, or about using a South-to-North ordering. More radically, one could experiment with orderings based purely on the socio-economic characteristics of each region, without taking spatial distances into account. While there might be orderings that lead to a significantly better forecasting performance than that achieved here under the proposed North-to-South ordering, performing a systematic assessment of the effect of ordering is infeasible under the complex model considered here, due to the substantial cost of model selection and fitting. 

Given that the search for an `optimal' ordering is impracticable, we design a semi-automatic model section procedure that does not explicitly take the ordering into account. In fact, while \cite{pourahmadi1999} proposed highly structured models for $\bold D_i$ and $\bold T_i$ which rely on the interpretation of their elements, and thus on their ordering, we use gradient boosting to choose which matrix elements should be modelled, and the effects that should be used to do so. The proposed approach is related to the method of \cite{stromer2022deselection}  
who use non-cyclical component-wise gradient boosting \citep{thomas2018gradient} to determine the effects' importance, 
and then run a further boosting procedure based on a subset of selected effects, chosen using
a user-defined importance threshold. In contrast, 
we use the out-of-sample predictive performance to determine the number of effects to include in the final model and we fit the latter using the methods from Section \ref{sec:model_fitting}, rather than boosting. 

In the following we describe the proposed model selection approach, and we refer to SM \ref{AppB1_PLS} for further details. 
For $j = 1, \dots, d$, we fit a univariate Gaussian GAM, $y_{ij} \sim N(\mu_{ij}, \sigma_{j}^2)$, using net-demand data from the 1st of January 2014 to the 31st of December 2016, with $\mu_{ij} = \eta_{ij}$ modelled via (\ref{mean_formula}). 
Then, define $\epsilon_{ij} = y_{ij} - \eta_{ij}$ and let $\bar{\bsS}$ be the empirical covariance matrix of such residuals. Having fixed $\bsEta_j$, for $j=1, \dots, d$,  gradient boosting spans the candidate effects appearing in 
\begin{align}\label{mcd_formula}
\eta_{ij} &= \bar{\eta}_{ij} + g_{j1}(t_i) + g_{j2}(t_i^2) + g_{j3}(\text{dow}_{i}) + f_{j1}^{10}(\text{doy}_i) + f_{j2}^{10}(\text{tod}_i) + \nonumber \\  & + \text{wcap}_i \times f_{j3}^{5}(\text{wsp}_{il_j}^{100}) + f_{j4}^{5}(\text{irr}_{il_j}) + f_{j5}^{5}(\text{temp}_{il_j}) + f_{j6}^{5}(\text{rain}_{il_j}^{1/2}) + f_{j7}^{5}(\text{n2ex}_{i})\,\,,
\end{align}
for $j=d+1, \dots, q$, where we indicate with $\bar{\eta}_{ij}$ the elements of the MCD of $\bar{\bsS}^{-1}$, which serve as fixed offsets, with $l_j$ the row of $\bold D_i$ or $\bold T_i$ on which the $j$-th linear predictor appears, and with $\text{wsp}_{il_j}^{100}$, $\text{irr}_{il_j}$ and so on the weather forecasts corresponding to the $l_j$-th region. Thus, we exploit the interpretation 
of the MCD parametrisation to specify the candidate weather forecasts modelling the non-trivial elements of the $l_j$-th row of $\bold D_i$ and $\bold T_i$. Indeed, the forecasts for the North of Scotland are used to model only $(\bold D_i)_{11}$, 
those for the South of Scotland are used to model $(\bold D_i)_{22}$ and second row of $\bold T_i$, and so on. To see the reasoning behind this choice, recall from Section  \ref{sec:mod_struct} that the linear predictors appearing on the $l_j$-row of $\bold T_i$ or $\bold D_i$ are related to, respectively, the coefficients or the residual variance of the regression of the $l_j$-th element of $\boldsymbol y$ on its predecessors. Hence, it seems reasonable to use the weather forecasts for the $l_j$-th region to model the effect of the preceding regions on $l_j$.

Note that (\ref{mcd_formula}) contains only a subset of the effects appearing in (\ref{mean_formula}). In particular, no bivariate tensor-product smooth effect is used to model $\bsEta_j$, $j=d+1, \ldots,q$. This choice is motivated by the fact that we are performing model selection across 105 linear predictors, so it is important to limit the number of candidate effects to ensure computational feasibility and statistical parsimony. For the same reason, the number of basis functions used to construct the effects in (\ref{mcd_formula}) is kept low. Further, the two terms $g_{j1}$ and $g_{j2}$ effectively form a single candidate effect in the model selection process.

To select the model for $\bsEta_{j}$, with $j=d+1, \dots, q$, we first run gradient boosting for $M = 3000$ iterations on the 2014-2016 data. At each iteration, the linear predictors fit the training data slightly better, which eventually leads to over-fitting. Having verified that over-fitting starts well before 3000 steps, we find the iteration $M^* \in \{1, 2, \dots , 3000\}$ at which the out-of-sample performance on 2017 data is optimal. The output of gradient boosting at step $M^*$ is 
a list containing the selected effect-linear predictor pairs and the cumulative log-likelihood gains obtained by adding them to the boosting model (see SM \ref{AppB1_PLS} for details). Assuming that the priority with which an effect-linear predictor pair should be added to the final model is proportional to its cumulative log-likelihood gain, the $L$ pairs corresponding to the $L$-largest gains should be included in the final model, for some $L\geq0$. 
Let $L_1 = 0, L_2 = 5, L_3 = 10, \dots$, 
be a grid of potential values for the total number of effects, $L$. To determine $L$, we optimise the predictive performance of the full multivariate Gaussian model on 2017 data. 
This is done by first fitting univariate Gaussian GAMs and adopting a 1-month block rolling origin forecasting procedure starting from the 1st of January 2017 to predict the value of $\eta_{ij}$, for $j=1, \dots, d$, covering the whole of 2017. Then, by using the same rolling procedure,  for each candidate value of $L_j$ we fit the multivariate Gaussian model $\bold y_i \sim N(\bsMu_{i}, \bsS_i)$ on 2017 data via the methods from Section \ref{sec:model_fitting}, obtaining the out-of-sample predictions for $\eta_{ij}$, $j=d+1,\ldots,q$, and the day-ahead predictions for $\bsMu_{i}$ and $\bsS_i$ are used  to compute the out-of-sample log-likelihood.
The procedure suggests including $L = 80$ effects. 
Running gradient boosting for 3000 steps and evaluating the predictive performance take, respectively, around 42 and 5 hours 
when run in parallel on a workstation with a 12-core Intel Xeon Gold 6130 2.10GHz CPU and 256GBytes of RAM.

\subsection{Model Selection Results} \label{sec:select_results}

\begin{figure}[t!]
\centering
\includegraphics[scale=0.7]{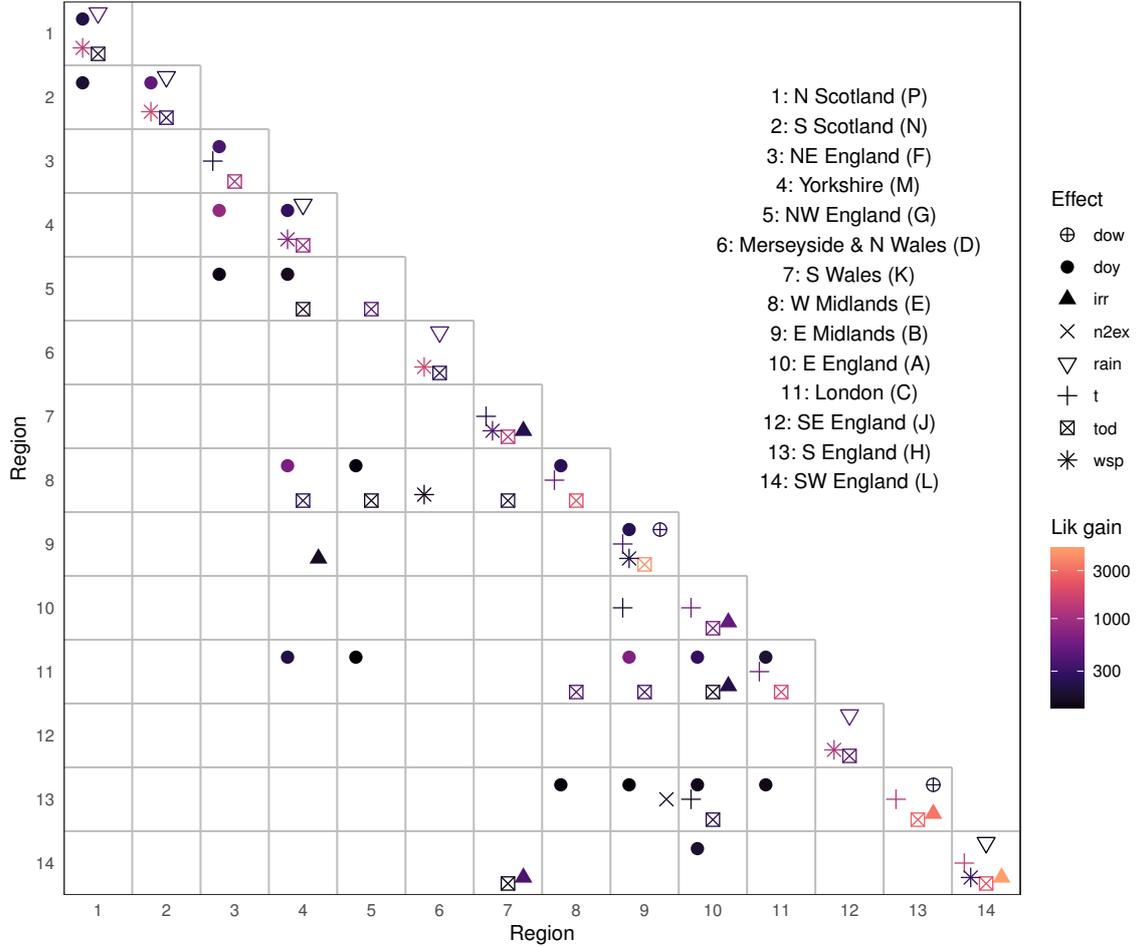} 
\caption{Model selection results. The diagonal corresponds to the elements of $\bold D_i$, the rest to those of $\bold T_i$. The symbols represent different effects and their colour is determined by the cumulative log-likelihood gain achieved by each effect during boosting. The elements of $\bold T_i$ corresponding to the empty cells are not zero, but modelled only via intercepts.}
\label{fig:sel_grid}
\end{figure}
\begin{figure}
\centering
\includegraphics[scale=0.24]{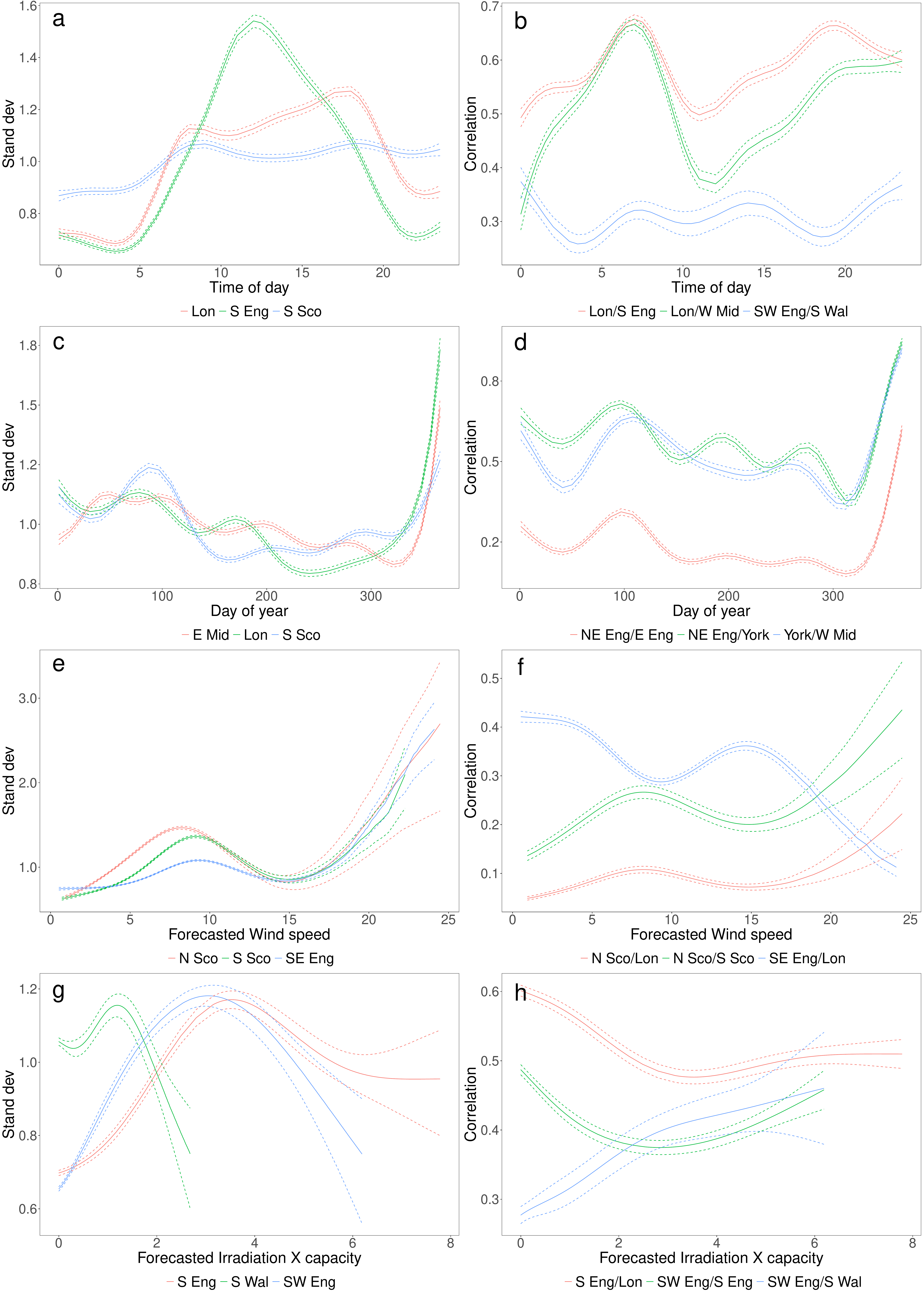} 
\caption{Left: ALEs of the time of day (a), day of year (c), forecasted wind speed (e) and solar irradiation (g) on the 
standard deviation of net-demand 
in a selected
group of regions. Right: ALEs of the same covariates on a selected group of pairwise correlations.}
\label{fig:ale_1}
\end{figure}

Figure \ref{fig:sel_grid} shows the effects selected to model each element of $\bold D_i$ and $\bold T_i$. Recall that, under the interpretation detailed in Section \ref{sec:mod_struct}, the elements on the $l$-th row of $\bold T_i$ are the coefficients of the regression of $y_{il}$ on $y_{il-1}$, $\dots$ $y_{i1}$, while the $l$-th diagonal element of $\bold D_i$ is the corresponding residual variance. Hence, the effects acting on $\bold D_i$ are not directly modelling the regional variances (i.e., the diagonal elements of $\bsS_i$), but the residual variance of the net-demand in each region, after having conditioned on the preceding regions. Similarly, the effects acting on $\bold T_i$ modulate the dependence between regions but they do not directly control correlations, which depend on the elements of both $\bold D_i$ and $\bold T_i$.
The effects of several covariates on regional variances and correlations are shown in Figure \ref{fig:ale_1} which contains a set of ALE plots, obtained by fitting a model containing the effects shown in Figure \ref{fig:sel_grid} to all the data available (2014-2018). Note that the 95\% credible bands showed in Figure \ref{fig:ale_1} are based on the Gaussian posterior approximation described in Section \ref{sec:inference}, which does not take into account the variability induced by the model selection procedure used here. Hence, the intervals might have coverage levels lower than the nominal ones.


Considering Figure \ref{fig:sel_grid}, note that most of the effects act on the diagonal elements of $\bold D_i$ and that the time of day, $\text{tod}_i$, affects all such elements. It is not surprising to see that the cumulative log-likelihood gain of the $\text{tod}_i$ effect is particularly large in highly urbanised areas, such as the Midlands (R. 8 and 9) and London (R. 11). The red ALE in Figure \ref{fig:ale_1}a shows the effect of daily seasonality in London, which is characterised by high net-demand variance during peak hours. The same effect has a similar, but flatter, shape in the South of Scotland (R. 2). In the South of England (R. 13) the effect has a single peak and it is even stronger than in London. As we discuss later, this is likely related to the high capacity of embedded solar generation relative to electricity consumption.

The time of day is used to model also several elements of $\bold T_i$. The strongest such effect, in terms of cumulative log-likelihood gain, acts on the element corresponding to London and the West Midlands  (R. 11 and 8). The ALE of $\text{tod}_i$ on the correlation between these regions is shown in green in Figure \ref{fig:ale_1}b. It shows that prediction errors in these urbanised regions are more correlated during the morning and evening ramps than in the middle of the day. The red curve shows that the correlation between London and the South of England (R. 13) has a similar pattern, although with milder daily oscillations. 

The effect of the day of the year, $\text{doy}_i$, is used to model many elements of $\bold D_i$ and $\bold T_i$. It is not surprising to see this effect appearing on the 8th and 11th row of Figure \ref{fig:sel_grid}, which correspond to the highly urbanised West Midlands (R. 8) and London (R. 11). As the green ALE in Figure \ref{fig:ale_1}c shows, net-demand forecast uncertainty is very high in London at the end of the year, due to holidays that have a sizeable, hard-to-model effect on demand patterns. Furthermore, as the effects in Figure \ref{fig:ale_1}d show, the uncertainty between regions is also highly correlated during this period, meaning that forecast errors are likely to have the same sign across regions as they are driven by the same underlying behavioural effects.

In accordance with intuition, wind speed, $\text{wsp}_{ij}^{100}$, is selected to model the elements of $\bold D_i$ corresponding to regions with a high penetration of embedded wind generation, such as the South East of England (R. 12), and the North (R. 1) and South (R. 2) of Scotland. The ALEs of $\text{wsp}_{ij}^{100}$ in these regions are shown in Figure \ref{fig:ale_1}e and could be interpreted as follows. At low wind, the variability of wind production is low because little or no generation is occurring, but it increases at modest wind speeds that are sufficient for power generation to occur, while being in a range where generation is highly sensitive to wind speed. Then variability decreases for high wind speeds, where power production is less variable as turbines self-regulate to maintain their maximum power production. At very high wind speeds, wind turbines may automatically shut down, and small differences in wind speed may result in large differences in production, leading to greater forecast uncertainty. 

It is perhaps surprising that wind speed is not selected to control the element of $\bold T_i$ controlling the dependency between the two Scottish GSPs, which both contain large amounts of embedded wind. However, capacity as a fraction of peak load is considerably higher in the North than in the South of Scotland. Further, the fact that $(\bold T_i)_{21}$ does not depend on $\text{wsp}_{ij}^{100}$, does not mean that the correlation between the two Scottish regions stays constant as $\text{wsp}_{ij}^{100}$ changes, as illustrated by the green curve in Figure \ref{fig:ale_1}f. The plot shows that the correlation is proportional to the variance in these regions, hence wind speed controls both the size and correlation between prediction errors. Interestingly, the blue curve in Figure \ref{fig:ale_1}f shows that the net-demand in the South East of England is less correlated with that in London (R. 11) as wind speed in the former region increases, which suggests that this covariate affects the net-demand patterns of these two regions quite differently.

The time of day and solar irradiance, $\text{irr}_{ij}$, are both 
strongly related to solar energy production, hence it is interesting to see that the effects of both variables are selected to model the elements of $\bold D_i$ corresponding to several Southern regions, which have high embedded solar generation capacity. The ALEs of $\text{irr}_{ij}$ on the net-demand variability in South Wales (R. 7), South (R. 13) and South-West (R. 14) England are shown in Figure \ref{fig:ale_1}g. Note that the horizontal scales are different because the installed solar capacity differs between regions. The shape of these effects is similar and could be interpreted as follows. Variability is low at low or high levels of irradiance which correspond to, respectively, heavily clouded (or night) and clear sky conditions. Variability is highest at intermediate levels of irradiance, which might correspond to partial or broken cloud conditions. However, the shape of the effects may also be affected by the correlation between irradiance and temperature and by changes in installed solar capacity over the study period, hence is it important not to over-interpret them. Solar irradiance is selected to control the dependency between the South of Wales and the South-West of England via the corresponding element of $\bold T_i$. This is interesting because, while the two regions are separated by the Bristol channel, they are geographically close, hence likely to be affected by similar weather patterns, and they both feature very high solar penetration relative to peak load.

As explained above, the set of effects selected by the semi-automated procedure proposed in Section \ref{sec:model_selection} matches intuition in many respects. However, looking at Figure \ref{fig:sel_grid}, note that more than half of the selected effects are related to calendar variables, namely progressive time, time of day, day of year and day of the week. Further, external temperature has not been selected to model any element of $\bold T_i$ or $\bold D_i$. Hence, it is interesting to analyse how the predictive performance of the model depends on the set of candidate covariates that are considered by the model selection procedure. In Sections \ref{sec:GSP14_eval} and \ref{sec:GSP14_eval_group} we analyse this issue and we compare the proposed model with two non-Gaussian alternatives.

\subsection{Validation on Regional Net-Demand Forecasting} \label{sec:GSP14_eval}

Here we assess the predictive performance of several alternative models obtained via the selection procedure proposed in Section \ref{sec:model_selection}. In particular, we consider three sets of candidate effects. The first set (\texttt{Full}) includes all the effects appearing in (\ref{mcd_formula}), hence it corresponds to the model analysed in Section \ref{sec:select_results}. Then we consider a calendar-only model (\texttt{Cal}), obtained by using only the first five effects in (\ref{mcd_formula}), that is from $g_{j1}(t_i)$ to $f_{j2}^{10}(\text{tod}_i)$, and a larger model (\texttt{Cal+Ren}) which also includes the effects of wind speed and solar irradiance. 
For each set of candidate effects, we use data from 2014 to 2017 to perform model selection, as described in Section \ref{sec:model_selection}.
Most of the selected effects in Figure \ref{fig:sel_grid} appear on the diagonal, hence we consider also a \texttt{Cal+Ren Diag} model, which is based on the same candidate effects as \texttt{Cal+Ren}, but where gradient boosting is allowed to model only the diagonal elements of ${\bold D}$,  while $\bold T$ is kept constant. 
See SM \ref{AppBX_MVN_models} for more details on the models discussed above.

Having selected the model structure, we assess performance on 2018 data. This is done by first fitting each model to data up to the end of 2017 and then forecasting net-demand during January 2018. We then refit the models to data up to the 31st of January 2018 and forecast net-demand for February. By iterating this rolling forecasting origin procedure, we obtain day-ahead predictions covering the whole of 2018, for each model. To speed up computation, we first fit model (\ref{mean_formula}) to the net-demand from each region using separate univariate Gaussian GAMs and then fit the corresponding residuals vectors using each of the covariance matrix models described above, via the methods described in Section \ref{sec:model_fitting}. 

We also include in the comparison three models where the marginal distribution of each GSP region's net-demand is modelled with an increasing amount of flexibility. Two of these models are useful to check whether relaxing the Gaussian assumption improves predictions.  In the \texttt{gaulss+cop} model, the net-demand of each region is modelled separately via a univariate location-scale Gaussian model.  In \texttt{shash+cop} the net-demand of each region is modelled via the four-parameter sinh-arcsinh distribution of \cite{jones2009sinh}, which nests the Gaussian but allows for asymmetry and fat-tails. The \texttt{shash+gpd+cop} model produces the same predictions of \texttt{shash+cop} between quantiles 0.05 and 0.95, but uses a generalised Pareto distribution (GPD) beyond these. Having fitted the univariate models separately to each GSP group, we evaluate the corresponding conditional c.d.f.s to obtain uniform residuals. Then we use a static Gaussian copula to model the correlation structure of the resulting 14-dimensional residual vectors. See SM \ref{AppB2_copula_models} for more details.

As for the multivariate Gaussian models, the Gaussian, sinh-arcsinh and GPD models are fitted to the raw residuals (responses minus estimated mean) of Gaussian GAMs based on formula (\ref{mean_formula}), hence their location parameters are kept constant to avoid fitting a location-like parameter twice. The effects used to model the scale parameters of the \texttt{gaulss+cop} model are chosen from the same pool of candidates used for \texttt{Cal+Ren}, following the approach described in Section \ref{sec:model_selection} but with the following modifications.  We set all the elements of $\bold T$ to zero which implies that the diagonal elements of $\bold D$ are directly controlling the marginal variance of each GSP group's net-demand (see Section \ref{sec:mod_struct}). Then, as for \texttt{Cal+Ren Diag}, we allow gradient boosting to model only the elements of $\bold D$. In this way, the models for the Gaussian scale parameter are selected to optimise the marginal fit to each region's net-demand. The resulting effects are used to model the scale parameters of the sinh-arcsinh distribution as well, while the parameters controlling the skewness and kurtosis are modelled only via intercepts. This is because our attempts to manually select a model for them led to a worse performance than what is reported below. Each GPD model is fitted to only 5\% of the data, hence we model its scale parameter using only a smooth effect of $\text{tod}_i$, while the shape parameter is kept constant. 

\begin{table}
\centering
\begin{tabular}{rrrrrrrr}    \hline Model  & Log & Log Ind & CRPS & Pin 001 & Pin 999 & Var 0.5 & Var 1.0 \\    \hline \texttt{Cal} & -3669 & -326.8 & 2582 & 21.67 & 44.51 & 9515 & 10293 \\    \texttt{Cal+Ren} & \underline{-4082} & -744.2 & 2571 & 20.24 & 33.97 & 9298 & 10012 \\    \texttt{Full} & -4071 & -771.9 & \underline{2570} & \underline{19.25} & 34.07 & \underline{9263} & \underline{9965} \\    \texttt{Cal+Ren Diag} & -3843 & -606.2 & 2574 & 20.97 & 35.45 & 9299 & 10014 \\    \texttt{gaulss+cop} & -3786 & -646.0 & 2576 & 21.05 & 32.07 & 9387 & 10125 \\    \texttt{shash+cop} & -3920 & -790.7 & 2574 & 20.66 & \underline{30.31} & 9355 & 10097 \\   \texttt{shash+gpd+cop} & -3975 & \underline{-821.6} & 2575 & 20.41 & 30.59 & 9362 & 10116 \\ \hline \end{tabular} 
\caption{Performance scores on 2018 test data, when forecasting the joint distribution of net-demand across the 14 GSP groups. The best score in each column is {\underline{underlined}}.}
\label{tab:GSP14_scores_new_GSP}
\end{table}

We use the day-ahead multivariate predictions of each model to compute the performance metrics reported in Table \ref{tab:GSP14_scores_new_GSP}. We consider the log score (i.e., the negative log-likelihood), the log score under independence (i.e., the sum of the negative marginal log-likelihoods of each GSP group), the marginal continuous ranked probability score (CRPS) and marginal pinball losses for quantiles 0.001 and 0.999 (each summed over the GSP groups, see \cite{gneiting2007} for a detailed introduction to both losses), and the $p$-variogram score \citep{scheuerer2015variogram} with $p=0.5$ and $p=1$.

\begin{figure}[htbp]
\centering
\includegraphics[scale=0.125]{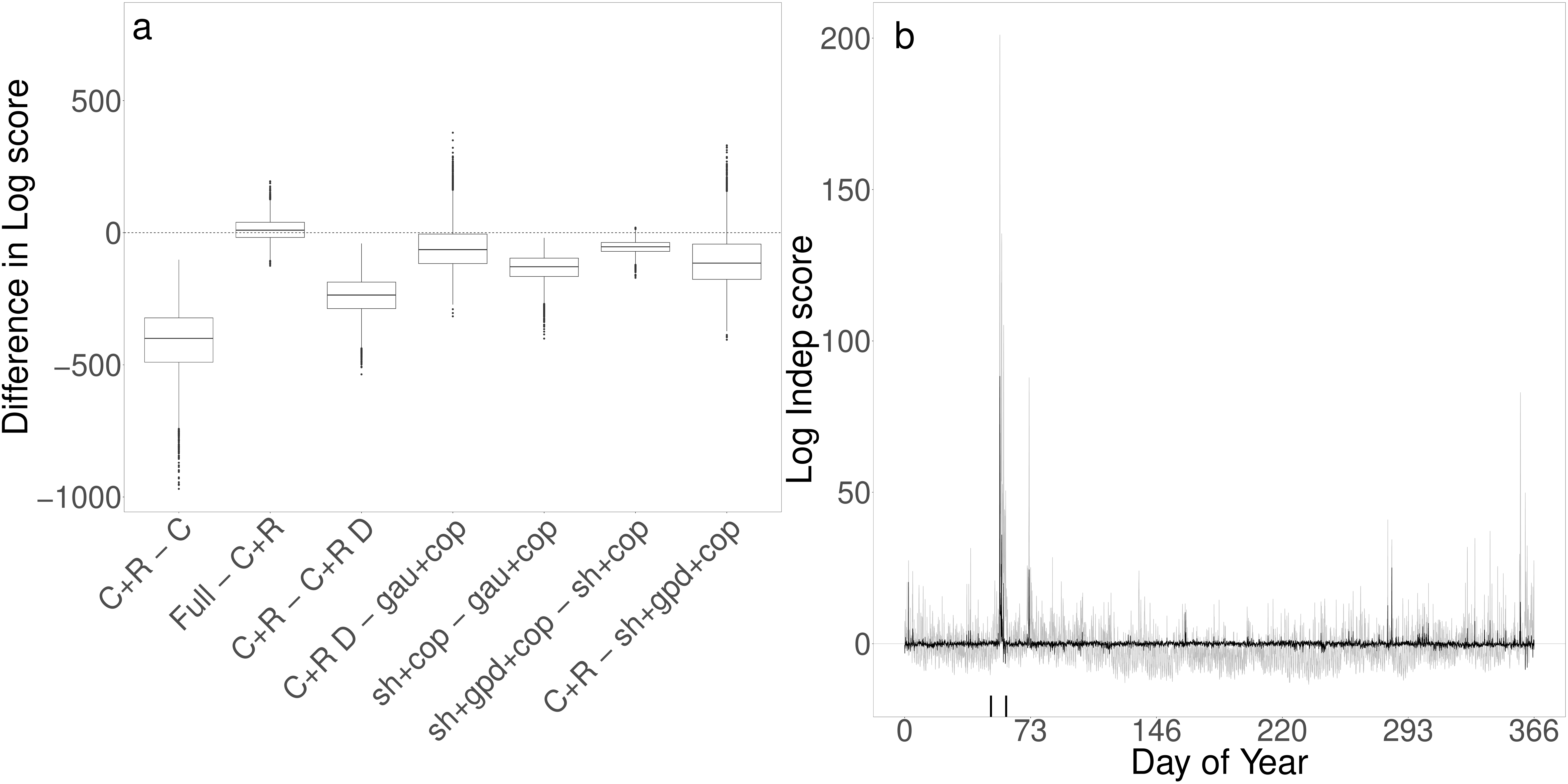}
\caption{a: Bootstrapped differences in the total log score between several pairs of models. Negative values mean that the first method is better than the second (e.g., \texttt{Cal+Ren} does better than \texttt{Cal}). b: Half-hourly marginal log losses of \texttt{Cal+Ren} (grey) along the test set (2018) and differences between the loss of \texttt{Cal+Ren} and that of  \texttt{shash+gpd+cop} model (black). The black ticks mark the start and the end of the Beast from the East cold wave.}
\label{fig:skills_plot}
\end{figure}

Considering Table \ref{tab:GSP14_scores_new_GSP}, note that the predictive performance of \texttt{Cal+Ren} is superior to that of \texttt{Cal} on all scores, demonstrating the importance of including covariates that are strongly related to embedded renewable generation. The significance of the improvement is demonstrated by Figure \ref{fig:skills_plot}a. Here, non-parametric bootstrapping with week-long blocks is used to quantify the variability of the differences in log scores between several pairs of models (SM \ref{AppB1_add_results} provides analogous plots for the remaining scores). The boxplots show that the gain obtained by including the effect of wind and solar irradiation is significant. Instead, extending the set of candidate effects as done in \texttt{Full} does not lead to any gain under the log score, which motivates our choice of basing the copula models on the same pool of candidate effects used for  \texttt{Cal+Ren}.

Table \ref{tab:GSP14_scores_new_GSP} and Figure \ref{fig:skills_plot}a show that modelling only the elements of $\bold D$ as done by \texttt{Cal+Ren Diag} leads to a worse performance, which highlights the importance of modelling $\bold T$ as well. Interestingly, \texttt{gaulss+cop} does worse than \texttt{Cal+Ren Diag} under the total log score, but better on the independent log score. As both models are Gaussian, this suggests that modelling the marginal variances directly and assuming that the correlation structure is constant, as done by \texttt{gaulss+cop}, leads to better marginal predictions but to a worse multivariate fit, relative modelling the $\bold D$ factor of the MCD parametrisation, as done by \texttt{Cal+Ren Diag} (recall that $\bold D$ affects the marginal variances as well as the correlation structure). Similarly, the \texttt{shash+cop} and \texttt{shash+gpd+cop} models have better marginal log scores than \texttt{Cal+Ren} (which uses the same pool of covariates), but worse total log and variogram scores. 
The fact that the CRPS loss, which is insensitive to tail behaviour 
\citep{taillardat2023evaluating}, is better under the \texttt{Cal+Ren} model suggests that this model is better at predicting intermediate marginal net-demand quantiles while, as demonstrated also by the pinball scores, the non-Gaussian marginal models do better in the upper tail.

To verify this, Figure \ref{fig:skills_plot}b shows the independent log score of the \texttt{Cal+Ren} model (grey line) and the differences between the \texttt{Cal+Ren} and \texttt{shash+gpd+cop} under the same score (black line). The largest gain of \texttt{shash+gpd+cop} relative to \texttt{Cal+Ren} occurs during the Beast from the East cold wave, which is delimited by the black ticks at the bottom of Figure \ref{fig:skills_plot}b. All the models considered here struggle to forecast the effect of this weather extreme on net-demand, as the training data does not contain any cold wave of similar magnitude. The better performance of \texttt{shash+gpd+cop} under the independent log score is explained by the fact that this model has fatter tails (recall that the tails are modelled separately from the bulk of the net-demand distribution under this model). To check to what degree the results discussed here are affected by the Beast from the East, in SM \ref{AppB1_add_results} we report a version of Table \ref{tab:GSP14_scores_new_GSP} obtained by excluding this exceptional period from the data. All the results discussed here are still valid but, as expected, excluding the cold wave reduces the benefit of adopting non-Gaussian marginal models.

\subsection{Validation on Macro-Regional Net-Demand Forecasting} \label{sec:GSP14_eval_group}

This work is motivated by the need for spatially coherent probabilistic net-demand forecasts in power flow studies and, as explained in Section \ref{chap2:intro}, the transmission grid boundaries of interest in such studies vary depending on, for example, the status of the network. Hence, it is interesting to verify whether the results discussed above still hold when the forecast is post-processed to match the needs of an operationally relevant scenario. While considering realistic scenarios would require covering engineering aspects that are well beyond the scope of this work, in Section \ref{chap2:intro} we proposed aggregating the GSP regions into the five macro-regions shown in Figure \ref{fig:GSP_groups}, which were motivated by some of the boundaries of interest described in the 2021 National Grid's Ten Year Statement \citep{elect10year}.

\begin{table}[ht]
\centering
\begin{tabular}{rrrrrrrr}    \hline Model  & Log & Log Ind & CRPS & Pin 001 & Pin 999 & Var 0.5 & Var 1.0 \\    \hline \texttt{Cal} & 3236 & 4529.7 & 2008 & 14.84 & 42.77 & 2215 & 4871 \\    \texttt{Cal+Ren} & 3091 & 4351.1 & 2000 & 13.63 & 34.73 & 2181 & 4782 \\    \texttt{Full} & \underline{3056} & \underline{4337.9} & \underline{1999} & \underline{12.92} & \underline{34.58} & \underline{2170} & \underline{4756} \\    \texttt{Cal+Ren Diag} & 3226 & 4469.3 & 2004 & 14.22 & 37.77 & 2188 & 4802 \\    \texttt{gaulss+cop} & 3184 & 4384.8 & 2003 & 14.64 & 36.59 & 2183 & 4788 \\    \texttt{shash+cop} &  &  & 2002 & 14.26 & 35.40 & 2181 & 4785 \\    \texttt{shash+gpd+cop} &  &  & 2003 & 13.94 & 37.47 & 2183 & 4792 \\     \hline \end{tabular} 
\caption{Performance scores on 2018 test data, when forecasting the joint distribution of net-demand across the five GSP macro-regions. The best score in each column is {\underline{underlined}}.}
\label{tab:GSP14_scores_new_GSP_region}
\end{table}

Joint macro-regional net-demand forecasts are easy to obtain because they are linear transformations of the regional forecasts. Table \ref{tab:GSP14_scores_new_GSP_region} 
shows the performance of each model when forecasting the joint distribution of macro-regional net-demand. Note that the scores are on a different scale due to the aggregation of net-demand (5 macro-regions vs 14 regions) and that the log scores have not been computed under two of the models, because the p.d.f. of linear combinations of correlated sinh-arsinh-distributed random variables is, to our best knowledge, not analytically available. The results are similar to those obtained at regional level, but here the \texttt{Full} model produces the best forecasts under all the scores. Due to the high cost of operating power systems (and the volume of energy traded in wholesale markets), even marginal improvements in forecast performance and associated decision-making can yield substantial economic and operational benefits.

It is also possible to linearly transform the joint regional forecasts to obtain marginal probabilistic forecasts of differences in net-demand between regions or macro-regions. Such forecasts could be of particular interest in the context of power flow analyses focused on specific transmission grid boundaries. In SM \ref{AppB1_add_results} we assess the performance of the models under three operationally motivated scenarios.  

\section{Conclusion} \label{sec:conclusions}

Forecasts of supply and demand are essential inputs to predict and manage power flows on electricity networks, as well as prices and other important variables. Given the imperative to maintain a reliable electricity supply, these predictions must enable risk to be quantified and managed. As the complexity of energy systems increases, the heuristic approaches widely used today are becoming inadequate and will have to be replaced by explicit probabilistic forecasts of power flows \citep{morales2013}.

Motivated by the need for spatially coherent, probabilistic net-demand forecasts to support energy system operations, we have focused on joint day-ahead forecasting of net-demand across the GSP regions comprising GB's transmission system. To accommodate for the dynamic nature of the net-demand covariance matrix, we let the elements of its MCD parametrisation vary with a number of temporal and weather-related covariates. To perform effect selection for a model comprising more than one hundred linear predictors, we leverage the interpretability of the chosen parametrisation and we combine it with a semi-automatic effect selection method, based on gradient-boosting. The results on the test set show that additive covariance matrix models significantly outperform, in terms of the total log, CRPS and variogram scores, two non-Gaussian models where the correlation matrix is static. However, the non-Gaussian models provide better predictions of extremely high quantiles, which suggests that a promising direction for future research might be adapting dynamic covariance matrix models to a non-Gaussian context. 

A further direction for future work would be to extend the model presented here to capture temporal, in addition to spatial, dependencies. In particular, the covariance matrix models used here implicitly assume that regional net-demand residual vectors are uncorrelated in time. While the mean vector model (\ref{mean_formula}) contains the effect of lagged net-demand, which is meant to capture part of the intra-regional temporal dependencies, more complex temporal effects could be captured by extending the covariance matrix to explicitly model the longitudinal nature of the data considered here. Such an extension should lead to models able to generate multivariate net-demand trajectories that are coherent both in space and in time, thus supporting important operations (e.g. determining the schedules for power-generating units) that must consider both spatial and temporal constraints. 


 
\section*{Acknowledgements} 

MF was supported by {\'E}lectricit{\'e} de France, JB by the EPSRC (EP/R023484/1 and EP/R023484/2) and RB by the Departmental Strategic Plan (PSD) of the University of Udine, Department of Economics and Statistics (2022-2025).

%


\putbib
\end{bibunit}

\newpage
\appendix
\begin{bibunit}
\section*{Supplementary Material to ``Additive Covariance Matrix Models: Modelling Regional Electricity Net-Demand in Great Britain''}

\label{AppA}

\setcounter{page}{1}

\textbf{Note}: The numbering of the tables and figures shown below follows from the main text. For example, when we mention Figure 1 below, we refer to Figure 1 in the main text. The first figure in the Supplementary Material below is Figure \ref{boot_scores_all} because the main text contains \ref{fig:skills_plot} figures.

\section{Derivatives of the Log-Likelihood} \label{AppA1}

\subsection{Setting up the Notation}

Consider a scalar-valued function $f$ of the $n$-dimensional vectors $\bsEta_1, \dots, \bsEta_q$. We indicate with $f^{\bsEta_k}$ and $f^{\bsEta_k\bsEta_j}$  the vectors with
$i$-th elements  
$$
f^{\eta_{ik}} = \frac{\partial f}{\partial \eta_{ik}} \quad \text{and} \quad
f^{\eta_{ik} \eta_{ij}} = \frac{\partial^2 f}{ \partial \eta_{ik} \partial \eta_{ij}}\,\,,
$$ 
where $\eta_{ik}$ indicates the $i$-th element of $\bsEta_k$.
Each $\bsEta_k$ is a function of a corresponding $p_k$-dimensional vector $\bsBeta_k$. For the derivatives of $f$ w.r.t. the elements of $\bsBeta_k$, we use the compact notation
$$
f^{\beta_{kr}} =  \frac{\partial f}{\partial \beta_{kr}} \quad \text{and} \quad
f^{\beta_{kr} \beta_{js}} = \frac{\partial^2 f}{ \partial \beta_{kr} \partial \beta_{js}}\,\,,
$$ 
where $\beta_{kr}$ indicates the $r$-th element of $\bsBeta_k$. Finally, we denote with $f^{\bsBeta_k} = \nabla_{\bsBeta_k} f$ the gradient of $f$ w.r.t. $\bsBeta_k$ and with $f^{\bsBeta_k\bsBeta_j} = \nabla_{\bsBeta_j}^\top \nabla_{\bsBeta_k} f$ the matrix of second derivatives.

\subsection{Gradient and Hessian w.r.t. $\bsBeta$}

To simplify the notation, let us indicate with $\bold y$ the collection of all response vectors $\bold y_1, \dots, \bold y_n$, and define $\mathcal L (\bsBeta) = \log p(\bsBeta|\bold y, \bsLambda)$. Recall that
\begin{equation*} \label{eq:log_posterior2}
\mathcal{L}(\bsBeta)  = \sum_{i=1}^n \ell_i  - \frac{1}{2} \bsBeta^\top \tilde{\bf S}^{\boldsymbol \lambda} \bsBeta
\end{equation*} 
where $\ell_i = \log p(\bold y_i|\bsBeta)$.
The gradient and Hessian of the log-posterior w.r.t $\bsBeta$ are
$$
    \mathcal L^{\bsBeta} (\bsBeta) = \sum_{i=1}^n \ell_i^{\bsBeta} - \tilde{\mathbf S}^{\bsLambda} \bsBeta \quad \text{and} \quad
    \mathcal L^{\bsBeta \bsBeta} (\bsBeta)
    =
    \sum_{i=1}^n \ell_i^{\bsBeta \bsBeta} - \tilde{\mathbf S}^{\bsLambda}\,\,.
$$

Let us define $\bar\ell = \sum_{i=1}^n \ell_i$ . To provide formulas for $\bar\ell^{\bsBeta}$ and $\bar\ell^{\bsBeta \bsBeta}$, let us assume that $\bsBeta = ({\bsBeta_1}^\top, \dots, {\bsBeta_q}^\top)^\top$, where $\bsBeta_j$ is the vector of regression coefficients specific to the $j$-th linear predictor, that is $\bsEta_{j} = \mathbf X^j \bsBeta_j$ where $\mathbf X^j$ is an $n \times p_j$ model matrix. With this notation, the $j$-th sub-vector of $\bar\ell^{\bsBeta}$ is 
$$
\bar\ell^{\bsBeta_j} = {\mathbf {X}^j}^\top \ell^{\bsEta_{j}}\,\,,
$$ 
while the $j, k$-th block of the Hessian is 
$$
\bar\ell^{\bsBeta_j \bsBeta_k} = (\bar\ell^{\bsBeta_k \bsBeta_j})^\top = {\mathbf {X}^k}^\top \text{diag}(\ell^{\bsEta_j\bsEta_k}) {\mathbf {X}^j}\,\,,
$$
where $\diag(\cdot)$ is the vector-to-matrix diagonal operator. The formulas provided so far apply to any GAM with multiple linear predictors and independent response vectors. In contrast, the expressions for $\ell^{\bsEta_j}$ and $\ell^{\bsEta_j\bsEta_k}$ are model-specific and are provided in the following section for a multivariate Gaussian distribution, with covariance matrix parametrised via the MCD.

\subsection{Derivatives w.r.t. $\bsEta$} \label{app:Dllk_Deta}

Let us start by defining a few useful quantities. Let $\bold G$ be a $(d-1)\times(d-1)$ lower triangular matrix such that $G_{jk}=C_{jk}+2d\mathbbm{1}_{\{k\leq j\}}$, where 
$$ {\rm C}_{jk}=\begin{cases}\quad \binom{j+1}{2}\quad \quad  \quad \,\,\,\,k=j \\  {\rm C}_{j(k+1)}-1\quad \quad \,  k<j\\  \quad\quad 0 \quad \quad \quad \quad \, \, k>j \,\, ,\end{cases}$$ 
 and $\mathbbm{1}$ is the indicator function. Define the $(d-1)\times(d-1)$ lower triangular matrices  $\bold Z$ and $\bold W$ such that ${\rm Z}_{jk}=k\mathbbm{1}_{\{k\leq j\}}$ and ${\rm W}_{jk}=(j+1)\mathbbm{1}_{\{k\leq j\}}$. Let $\bold z=\rvech( \bold Z) $ and  $\bold w=\rvech( \bold W)$, where $\rvech(\cdot)$ is the row-wise half-vectorisation operator, that is $\rvech(\bold Z) = (Z_{11}, Z_{21}, Z_{22}, Z_{31}, Z_{32}, Z_{33}, \dots, Z_{(d-1)(d-1)})^\top$. Let  $\bold{Q}_l$, for $l=1, \ldots, d$, and  $\bold{P}_l $, for $ l=1, \ldots, d(d-1)/2$,  be $d\times d$ matrices such that $(\bold{Q}_l)_{ll}=e^{-\eta_{l+d}}$ and $(\bold{P}_l)_{{z}_{l}{w}_{l}}=1$, while all other elements are equal to zero.

Here, index $i$ is not needed, hence we drop it and we indicate the $i$-th log-likelihood component $\ell_i$ simply with $\ell$. Note that, given that we are focusing on an individual $i$, here $\bsEta$ is a $q$-dimensional vector and $q = d + d(d+1)/2$. If we omit the constants that do not depend on $\bsEta$ and we indicate with $r_k$ the $k$-th element of the residual vector, $\bold r = \bold y- \boldsymbol \mu$, the Gaussian log-density can be written
\begin{align*}
\ell & = -\frac{1}{2} \left\{\tr(\log \bold D^2)+\bold r^\top\bold{T}^\top\bold{D}^{-2}\bold{T}\bold r\right\} \nonumber \\
& = -\frac{1}{2}\sum_{j=1}^{d}\Big\{\eta_{j+d} + e^{- \eta_{j+d}}\Big(\sum_{k=1}^{j-1} \eta_{{\rm G}_{(j-1)k}} r_k+ r_j \Big)^2\Big\}\,\, ,
\end{align*}
where we used $\log |\boldsymbol \Sigma| = \tr(\log \bold{D}^2)=\sum_{j=1}^{d}  \eta_{j+d}$ and we implicitly assumed that the sum $\sum_{k=1}^{j-1}$ should not be computed when $j = 1$ (we will use the same convention in several places below). Similarly, below we assume that $\sum_{j=l+1}^{d}$ will not be computed when $l=d$. Here we provide the first and second derivatives of $\ell$ w.r.t. ${\bsEta}$ both in compact matrix form and in an extended format, the latter being more useful for efficient numerical implementation.

With notation above, the elements of $\ell^{\bsEta} = (\ell_1^{\bsEta}, \dots, \ell_q^{\bsEta})^\top = (\partial \ell/ \partial \eta_1, \dots, \partial \ell/ \partial \eta_q)^\top$ are 
\begin{align*}
\ell^{\bsEta}_l & =\left ( \bold{T}^\top\bold{D}^{-2}\bold{T}\bold{r} \right )_l   \\  
&=e^{- \eta_{d+l}}\Big(\sum_{k=1}^{l-1} \eta_{ {\rm G}_{(l-1)k}}  r_k +  r_l\Big) + \sum_{j=l+1}^{d}e^{-\eta_{j+d}}\Big(\sum_{k=1}^{j-1}  \eta_{ {\rm G}_{(j-1)k}} r_k+ r_j \Big) \eta_{ {\rm G}_{(j-1)l}} \,\, ,
\end{align*}
for $l=1, \ldots, d$,  %
\begin{align*}
\ell^{\bsEta}_l & =  \frac{1}{2}\bold{r}^\top \bold{T}^\top\bold{Q}_{l-d}\bold{T}\bold{r} -\frac{1}{2} \\  
& =\frac{1}{2}e^{-\eta_l}\Big(\sum_{k=1}^{l-d-1} \eta_{ {\rm G}_{(l-d-1)k}} r_k+ r_{l-d} \Big)^2-\frac{1}{2}\,\, ,
\end{align*}
for $l=d+1, \ldots, 2d$, and 
\begin{align*}
\ell^{\bsEta}_l &= -\bold{r}^\top \bold{P}_{l-2d}\bold{D}^{-2}\bold{T}\bold{r}     \\
&= -e^{ \eta_{w_{l-2d}+d}}\Big(\sum_{k=1}^{w_{l-2d}-1} \eta_{ {\rm G}_{( w_{l-2d}-1)k}} r_k+ r_{ w_{l-2d}} \Big) r_{ z_{l-2d}}\,\,,  
\end{align*}
for $l=2d+1, \ldots, q$. 

The elements forming the upper triangle of $\ell^{\bsEta\bsEta}$ (here $\ell^{\bsEta\bsEta}_{lm} = \partial^2 \ell / \partial \eta_l \partial \eta_m$), are 
\begin{align*}
\ell^{\bsEta\bsEta}_{lm} &=-\left ( \bold{T}^\top\bold{D}^{-2}\bold{T} \right )_{lm} \\
&=-\Big\{e^{-{\eta}_{l+d}} + \sum_{k=l+1}^{d}e^{-{\eta}_{k+d}} \Big({\eta}_{ {\rm G}_{(k-1)l}}  \Big)^2\Big\}\mathbbm{1}_{\{m=l\}} \\
&\hspace{5mm}-\Big(e^{-{\eta}_{m+d}} {\eta}_{ {\rm G}_{(m-1)l}}+\sum_{k=m+1}^{d}e^{-{\eta}_{k+d}} {\eta}_{ {\rm G}_{(k-1)l}}  {\eta}_{ {\rm G}_{(k-1)m}} \Big) \mathbbm{1}_{\{m>l\}}\,\,,
\end{align*}
for $l = 1, \ldots, d,$ and $m=l, \ldots, d$,

\begin{align*}
\ell^{\bsEta\bsEta}_{lm}&=-\left ( \bold{T}^\top\bold{Q}_{m-d}\bold{T}\bold{r} \right )_l  \\ 
&=-e^{-{\eta}_{m} }\Big\{\Big(\sum_{k=1}^{l-1}  \eta_{ G_{(l-1)k}} r_k+ r_{l} \Big) \mathbbm{1}_{\{m-d=l\}} \\ &\hspace{5mm}+\Big(\sum_{k=1}^{m-d-1}  \eta_{ G_{(m-d-1)k}} r_k+ r_{m-d} \Big) \eta_{ G_{(m-d-1)l}}  \mathbbm{1}_{\{m-d>l\}}\Big\}\,\, ,
\end{align*}
for $l=1, \ldots, d,$ and $m=d+1, \ldots, 2d$,  
\begin{align*}
\ell^{\bsEta\bsEta}_{lm}&=\left ( \bold{P}_{m-2d}\bold{D}^{-2}\bold{T}\bold{r}  + \bold{T}^\top\bold{D}^{-2}\bold{P}^\top_{m-2d} \bold{r}  \right )_l\,\, \\  
&=e^{- \eta_{w_{m-2d}+d}}\Big\{  r_{ z_{m-2d}}\Big(\mathbbm{1}_{\{ w_{m-2d}=l\}}+ \eta_{{\rm G}_{( w_{m-2d}-1)l}}\mathbbm{1}_{\{ w_{m-2d}>l\}}\Big) \\
&\hspace{5mm}+\Big(\sum_{k=1}^{ w_{m-2d}-1}  \eta_{ {\rm G}_{( w_{m-2d}-1)k}} r_k+r_{ w_{m-2d}} \Big)\mathbbm{1}_{\{ z_{m-2d}=l\}}\Big\}\,\, , 
\end{align*}
for $l=1, \ldots, d,$ and $m=2d+1, \ldots, q$, 
\begin{align*}
\ell^{\bsEta\bsEta}_{lm}&=-\frac{1}{2}\bold{r}^\top \bold{T}^\top\bold{Q}_{l-d}\bold{T}\bold{r}  \\
&=-\frac{1}{2}e^{- \eta_{l}}\Big(\sum_{k=1}^{l-d-1} \eta_{{\rm G}_{(l-d-1)k}}  r_{k}+ r_{l-d} \Big)^2\mathbbm{1}_{\{ m=l\}}\,\, ,  
\end{align*}
for $l = d+1, \ldots, 2d,$ and $m=l, \ldots, 2d$,
\begin{align*}
\ell^{\bsEta\bsEta}_{lm}&=\bold{r}^\top \bold{P}_{m-2d}\bold{Q}_{l-d}\bold{T}\bold{r}   \\
&=e^{- \eta_{l}}\Big(\sum_{k=1}^{l-d-1} \eta_{ {\rm G}_{(l-d-1)k}}  r_{k}+ r_{l-d} \Big)r_{ z_{m-2d}}\mathbbm{1}_{\{ w_{m-2d}=l-d\}}\,\, ,
\end{align*}
for $l=d+1, \ldots, 2d,$ and $m=2d+1, \ldots, q$, and finally
\begin{align*}
\ell^{\bsEta\bsEta}_{lm}&=-\bold{r}^\top \bold{P}_{l-2d}\bold{D}^{-2}\bold{P}^\top_{m-2d}\bold{r} \\
&=-e^{-{\eta}_{ w_{l-2d}+d}}  r_{ z_{l-2d}} r_{ z_{m-2d}} \mathbbm{1}_{\{ w_{m-2d}=w_{l-2d}\}} \,\, ,
\end{align*}
for $l = 2d+1, \ldots, q,$ and $m = l, \ldots, q$.

\subsection{Details on the Accumulated Local Effects} \label{app:jacobians}

Recall that $\bsS_i$ depends on the covariate vector $\boldsymbol x_i$ via the linear predictor vector $\bsEta_i$. By omitting the subscript $i$ and expliciting the dependence on $\boldsymbol x$, let us denote with $\omega(\boldsymbol x)$ 
a generic element of $\bsS$ or of the correlation matrix $\boldsymbol \Gamma$, the elements of the latter being defined by $\Gamma_{jk}=\Sigma_{jk}/\sqrt{\Sigma_{jj}\Sigma_{kk}}$. Assuming that $\omega(\boldsymbol x)$ is differentiable w.r.t. the $k$-th covariate, the main (first-order) accumulated local effects (ALEs) of $x_k$ is 
\begin{equation*} \label{eq:ALE_1}
\overbar\omega_{k}(x) = \int_{x_{k}^{\text{min}}}^{x} \mathbb{E}_{\bsx_{\backslash k}} \left \{ \omega^k(z, \bsx_{\backslash k})|x_k = z \right \} d z - \text{constant}\,\,,
\end{equation*}
where $\bsx_{\backslash k}$ is $\bsx$ with the $k$-th element excluded, $\omega^k = \partial \omega / \partial x_k$ and $\mathbb{E}_{\bsx_{\backslash k}}\{ \cdot | x_k = z \}$ is the conditional expectation w.r.t. $p(\bsx_{\backslash k}|x_k = z)$. The choice of $x_{k}^{\text{min}}$ is unimportant, as changing it simply shifts the effect vertically, so in practice $x_{k}^{\text{min}}$ is set to just below the smallest observed value of $x_k$. 

An uncentered first-order ALE is obtained by setting the constant term to zero, while a centered ALE has a mean equal to zero when averaged over the observed values of the covariate of interest. This is explained by \cite{apley2020}, who also provide formulas for obtaining estimated effects ${\hat{\overbar{\omega}}}_{k}(x)$, by approximating the integral above. For instance, consider an uncentered ALE and let $x_{ik}$ be the $i$-th observed value of $x_k$. Further, denote with $z_{0k}, \ldots, z_{Bk}$ a grid of values along $x_k$, with $z_{0k}=\min_{i=1, \ldots, n} x_{ik}$ and $z_{Bk}=\max_{i=1, \ldots, n} x_{ik}$, and let $n_{k}(1), \ldots, n_k(B)$  
the number of $x_{ik}$ included, respectively, in the intervals $N_k(1)=[z_{0k}, z_{1k}), \ldots, N_k(B)=[z_{B-1k}, z_{Bk}]$. Then, the ALE of $x_k$ is estimated via 
$${\hat{\overbar{\omega}}}_{k}(x)= \sum_{v=1}^{v_k(x)}\frac{1}{n_k(v)} \sum_{\{i: x_{ik} \in N_k(v)\}} \{\omega(z_{vk}, \boldsymbol x_{i \backslash k}) - \omega(z_{v-1k}, \boldsymbol x_{i \backslash k})\}  $$
with ${\hat{\overbar{\omega}}}_{k}(z_{0k})=0$ and $v_k(x) \in \{1, \ldots,B\}$  denoting the bin number to which an arbitrary value $x$ of $x_k$ belongs.

The uncertainty of ALEs can be quantified by propagating posterior parameter uncertainty via a standard asymptotic approximation. Recall that standard Bayesian asymptotics justify approximating $p(\bsBeta|\bold y, \bsLambda)$ with a Gaussian distribution, $N(\hat{\bsBeta}, {\bf V}_{\bsBeta})$, centered at the MAP estimate and with covariance matrix ${\bf V}_{\bsBeta} = -\boldsymbol {\mathcal{H}}^{-1}$, where $\boldsymbol {\mathcal{H}}$ is the negative Hessian of $\log p(\bsBeta|\bold y, \bsLambda)$, evaluated at $\hat \bsBeta$. 
\cite{capezza2021additive} show that, in a GAM context, the delta method can be used to approximate the posterior variance of ALEs  
 via $\text{var}\{{\hat{\overbar{\omega}}}_{k}(x)\} \approx \nabla_{\bsBeta}^\top{\hat{\overbar{\omega}}}_{k} {\bf V}_{\bsBeta} \nabla_{\bsBeta} {\hat{\overbar{\omega}}}_{k}$. The authors provide formulas for the Jacobian $\nabla_{\bsBeta} {\hat{\overbar{\omega}}}_{k}(x)$ that apply to any GAM with multiple linear predictors, covering also the case where $x_k$ is a categorical variable. Below we provide the details for obtaining the Jacobian $\nabla_{\bsBeta} {\hat{\overbar{\omega}}}_{k}(x)$, where the only output-specific component is the Jacobian of the parametrisation linking $\omega(\boldsymbol x)$ with  $\bsEta$.

Suppose that the output of interest, $\omega(\boldsymbol x)$, is an element of $\bsS$. Define the vector $\boldsymbol\sigma_{jk} = \{(\bsS_1)_{jk}, \dots, (\bsS_n)_{jk}\}^\top$ and consider the vector containing the values of the $j$-th linear predictor at each observation, that is ${\bsEta}_{j} = (\eta_{1j}, \dots, \eta_{nj})^\top$. For $j=1, \dots, q$, we have that $\bsEta_{j}={\bold X}^{j}\bsBeta_{j}$ where ${\bold X}^j$ and $\bsBeta_j$ are, respectively, the $n \times p_j$ model matrix and the $p_j$-dimensional vector of regression coefficients belonging to the $j$-th linear predictor. 
The, the $n \times p$ Jacobian matrix of $\boldsymbol\sigma_{jk}$ w.r.t. $\bsBeta$ is
\[ 
\mathbf{J}^{jk} = \nabla_{\bsBeta}^\top\boldsymbol \sigma_{jk} = ( \nabla_{\bsBeta_1}^\top \boldsymbol\sigma_{jk}, \cdots, \nabla_{\bsBeta_{q}}^\top \boldsymbol\sigma_{jk})\,\,,
\]
where $p = \sum_{j=1}^{q} p_j$. The $a$-th block of $\mathbf{J}^{jk}$ is 
$$
\nabla_{\bsBeta_a}^\top \boldsymbol\sigma_{jk} = \nabla_{\bsEta_{a}} ^\top \boldsymbol\sigma_{jk} \nabla^\top_{\bsBeta_a} {\bsEta_a} = \nabla_{\bsEta_{a}} ^\top \boldsymbol\sigma_{jk} {\bf X}^a\,\,,
$$
for $a = 1, \dots, q$, and where $\nabla_{\bsEta_{a}} ^\top \boldsymbol\sigma_{jk}$ is an $n \times n$ diagonal matrix with non-zero elements
$$
\left(\nabla_{\bsEta_{a}} ^\top \boldsymbol\sigma_{jk}\right)_{ii} = \frac{\partial (\bsS_i)_{jk}}{\partial \eta_{ia}}\,\,.
$$
Note $\partial (\bsS_i)_{jk}/\partial \eta_{ia}$ is the only parametrisation-dependent component of the Jacobian, thus in Section \ref{sec:D_Sig_D_Eta} we provide the relevant formulas. When $\omega(\boldsymbol x)$ is an element of $\bold \Gamma$, the posterior variance of the ALEs is approximated similarly, since the Jacobian of $\{(\boldsymbol \Gamma_1)_{jk}, \dots, (\boldsymbol \Gamma_n)_{jk}\}$ w.r.t. $\bsBeta$ is computed analogously, but with $\partial (\boldsymbol \Gamma_i)_{jk}/\partial \eta_{ia}$ in place of $\partial (\bsS_i)_{jk}/\partial \eta_{ia}$. Formulas for $\partial (\boldsymbol \Gamma_i)_{jk}/\partial \eta_{ia}$ are provided in Section \ref{sec:D_Corr_D_Eta}.

\subsubsection{Derivatives of $\boldsymbol{\Sigma}$ w.r.t. $\bsEta$} \label{sec:D_Sig_D_Eta}

Here the index $i$ is not needed, hence we drop it. Consider the factorisation $\boldsymbol{\Sigma} = \bold R \bold R^\top$ where $\bold R = \bold L \bold D$ and $\bold L= \bold T^{-1}$. The partial derivative of the $(l,m)$ element of $\boldsymbol{\Sigma}$ w.r.t. $\eta_j$, is
$$
\frac{\partial \Sigma_{lm}}{\partial \eta_j}= \sum_{k=1}^d \left( \frac{\partial {\rm R}_{lk}}{\partial \eta_j}{\rm R}_{mk} +  {\rm R}_{lk}\frac{\partial {\rm R}_{mk}}{\partial \eta_j} \right),
$$
where 
$$
\frac{\partial {\rm R}_{lk}}{\partial \eta_j} = 0 \,\,, \quad \text{for} \quad j=1, \ldots, d,
$$
$$
\frac{\partial {\rm R}_{lk}}{\partial \eta_j} = \frac{\partial {\rm L_{lk}} {\rm D_{kk}}}{\partial \eta_j} =\frac{1}{2} {\rm L_{l(j-d)}} {\rm D_{(j-d)(j-d)}} \mathbbm{1}_{\{ j-d=k\}}\,\,, \quad \text{for} \quad j=d+1, \ldots, 2d,
$$
and
$$
\frac{\partial {\rm R}_{lk}}{\partial \eta_j} = \frac{\partial {\rm L_{lk}} {\rm D_{kk}}}{\partial \eta_j}=- {\rm L_{ls}}  {\rm L_{tk}}  {\rm D_{kk}}\,\,, \quad \text{for} \quad j=2d+1, \ldots, q,
$$ 
with $t$ and $s$ being the row and column indices of the only element of $\bold T$ that depends on $\eta_j$, for $j=2d+1, \ldots, q$. Hence, we obtain
$$
\frac{\partial \Sigma_{lm}}{\partial \eta_j} = 0\,\,, \quad \text{for} \quad j=1, \ldots, d, 
$$
$$
\frac{\partial \Sigma_{lm}}{\partial \eta_j}  = {\rm L_{l(j-d)}}{\rm L_{m(j-d)}} {\rm D_{(j-d)(j-d)}}\,\,, \quad \text{for} \quad j=d+1, \ldots, 2d, 
$$
and
$$
\frac{\partial \Sigma_{lm}}{\partial \eta_j}=- {\rm L_{ls}} \sum_{k=1}^{d}{\rm L_{tk}} {\rm D^2_{kk}} {\rm L_{mk}} - {\rm L_{ms}} \sum_{k=1}^{d}{\rm L_{lk}} {\rm D^2_{kk}}{\rm L_{tk}}\,\,, \quad \text{for} \quad j=2d+1, \ldots, q,
$$
where $t$ and $s$ are defined as above.

\subsubsection{Derivatives of $\boldsymbol{\Gamma}$ w.r.t. $\bsEta$} \label{sec:D_Corr_D_Eta}

To simplify the notation indicate $(\Sigma_{ll})^{-1/2}$ with $\Sigma^{-1/2}_{ll}$. The $(l,m)$ element of $\boldsymbol{\Gamma}$ is  
$$
\Gamma_{lm}= \Sigma^{-1/2}_{ll} \, \Sigma_{lm} \, \Sigma^{-1/2}_{mm}\,\, ,
$$
and its partial derivative w.r.t. $\eta_j$, for $j = 1, \dots, q$, is
$$
\frac{\partial \Gamma_{lm}}{\partial \eta_j} =\Sigma^{-1/2}_{ll} \, \frac{\partial \Sigma_{lm}}{\partial \eta_j} \,
\Sigma^{-1/2}_{mm}-\frac{1}{2}\Sigma_{lm}\Bigg\{\Sigma^{-3/2}_{ll} \, \Sigma^{-1/2}_{mm} \, \frac{\partial \Sigma_{ll}}{\partial \eta_j} +\Sigma^{-1/2}_{ll} \, \Sigma^{-3/2}_{mm} \, \frac{\partial \Sigma_{mm}}{\partial \eta_j} \Bigg\}\,\,,
$$
 and where the derivatives of the elements of $\bsS$ w.r.t. $\eta_j$ are provided in Section \ref{sec:D_Sig_D_Eta}.

\label{AppB}

\section{Further Details and Results} \label{AppB1}

\subsection{Details on the Model Selection Approach} \label{AppB1_PLS}

In the following we provide more details on the model selection approach described in Section \ref{sec:model_selection}.

Denote with $\mathcal{R}_j$ a vector containing the indices of the candidate effects in (\ref{mcd_formula}), with the first element referring to both $g_{j1}$ and $g_{j2}$, so that these two terms effectively form a single effect in the model selection process.
Let $\bsEta_{j}$ be the $j$-th linear predictor and indicate with $\bsEta$ the $n \times q$ matrix containing all the linear predictors. 
Let $\bold X^{r}$, with $r \in \mathcal{R}_j$, be the model matrix of the $r$-th effect, and let ${\bf S}_r$ be the corresponding positive semi-definite penalty matrix. Let $\boldsymbol \Delta$ be a list of length $q$, its $j$-th element $\boldsymbol \Delta_j$ being a vector of dimension $\text{card}(\mathcal{R}_j)$ with all elements, $\Delta_{rj}$, initialised at 0.  

Algorithm \ref{algo:grad_boosting} details the steps of the gradient boosting procedure used to quantify the importance of each candidate effect-linear predictor pair. In particular, its output is $\boldsymbol \Delta$, the list containing the cumulative log-likelihood gains achieved by each candidate effect-linear predictor pair. Note that in step 1.II.(a) we regress the log-likelihood gradient $\bold u_j$ on the model matrix of each effect in $\mathcal{R}_j$ using penalised least squares, with penalty $\zeta_r{\bf S}_r$ where $\zeta_r > 0$. As explained by \cite{hofner2011framework}, penalisation is helpful to mitigate the selection bias in favour of effects with more parameters.  In particular, the penalties of each effect should be scaled to make sure that all the effects have similar effective degrees of freedom. The latter are defined as
$$
\text{edf}_r = \tr{\left\{\bold X^{r}({\bold X^{r}}^\top\bold X^{r} + \zeta_r{\bf S}_r)^{-1}{{\bold X}^{r}}^\top\right\}},
$$
and depend on $\zeta_r$. In particular, assuming that $\bold X^{r}$ is of full rank $p_r$ and that ${\bf S}_r$ has rank $s_r \leq p_r$, as $\zeta_r$ increases from zero to infinity the $\text{edf}_r$ decrease from $p_r$ to $p_r - s_r$.

For each effect appearing in the MCD model (\ref{mcd_formula}) such that $p_r > 4$, we choose $\zeta_r$ such that $\text{edf}_r = 4$. This is a one-dimensional numerical optimisation problem, which can be solved very rapidly prior to running Algorithm \ref{algo:grad_boosting}. The effect of progressive time $t_i$ is modelled using two parameters, hence penalisation is unnecessary.
The penalty matrices ${\bf S}_r$ correspond to cubic splines penalties (i.e., proportional to the integrated second derivative of the effect, $\int f''(x)^2dx$) for all the smooth effects in model (\ref{mcd_formula}), while a standard ridge penalty is used for the effect of the factor $\text{dow}_i$.

\begin{algorithm} \caption{Quantifying the effects' importance via gradient boosting}
\label{algo:grad_boosting}
\begin{algorithmic}[1]
\STATE  For $j = d+1, \dots, q,$
\begin{enumerate}
\item[I.] Compute the gradient $\bold u_j$ of the log-likelihood w.r.t. $\bsEta_j$, that is  
$$
u_{ij} = \frac{\partial \log p(\bold y_i|\bsEta_{i})}{\partial \eta_{ij}}\,\,, \quad \text{for} \quad i = 1, \dots, n,
$$
where $\bsEta_{i}$ is the $i$-th row of $\bsEta$.

\item[II.] For $r \in \mathcal{R}_j$ 

\begin{enumerate}

\item Regress $\bold u_j$ on $\bold X^{r}$ via 
$$
\hat{\bold u}_j^r = \bold X^{r}({\bold X^{r}}^\top\bold X^{r} + \zeta_r{\bf S}_r)^{-1}{{\bold X}^{r}}^\top \bold u_j. 
$$

\item Update the corresponding linear predictor via
$
\tilde{\bsEta}_j^r = \bsEta_j + \nu \hat{\bold u}_j^r
$, 
where $\nu = 0.1$ is the learning rate. Let $\tilde{\bsEta}^{rj}$ be the same as $\bsEta$ matrix, but with the $j$-th column set to $\tilde{\bsEta}_j^r$.

\item Compute the corresponding change in log-likelihood 
$$
\delta_{rj} = \sum_{i=1}^n \left \{ \log p(\bold y_i|\tilde{\bsEta}_i^{rj}) - \log p(\bold y_i|\bsEta_{i}) \right \}.
$$
\end{enumerate}

\end{enumerate}

\STATE Let $j^*$ and $r^*$ be the indices corresponding to the largest $\delta_{rj}$. Update the relevant linear predictor and the cumulative gain vector by doing
$$
\bsEta_j \leftarrow \tilde{\bsEta}_{j^*}^{r^*}, \quad \text{and} \quad
 \Delta_{r^*j^*} \leftarrow \Delta_{r^*j^*} + \delta_{r^*j^*}\,\,.
$$

\STATE Unless the maximum number of iterations $M$ has been reached, go back to step 2.

\end{algorithmic}
\end{algorithm}

\newpage 

\subsection{Details on the Multivariate Gaussian Models} \label{AppBX_MVN_models}

As explained in Section \ref{sec:GSP14_eval}, the \verb|Full|, \verb|Cal+Ren|, \verb|Cal| and \verb|Cal+Ren Diag| models are based on a conditional multivariate Gaussian distribution, parametrised via the MCD. The purpose of this section is to provide a self-contained introduction to these models.

Each element of the mean vector is modelled via formula (\ref{mean_formula}), which we repeat here for convenience
\begin{align*}
\eta_{ij} &= g_{j1}(t_i) + g_{j2}(t_i^2) + g_{j3}(\text{dow}^+_{ij}) + g_{j4}(\text{shol}_{ij}) +  g_{j5}(y_{ij}^{24}) + g_{j6}(\text{wsp}_{ij}^{10}) \nonumber \\ & + f_{j1}^{20}(\text{doy}_i) + f_{j2}^{35}(\text{tod}_i) + f_{j3}^{10}(\text{n2ex}_i) + f_{j4}^{35}(\text{temp}_{ij}) + f_{j5}^{35}(\text{temp}_{ij}^S) + f_{j6}^{10}(\text{rain}_{ij}^{1/2}) \nonumber \\ & + \text{wcap}_i \times f_{j7}^{20}(\text{wsp}_{ij}^{100}) + f_{j8}^{5}(\text{irr}_{ij}) + f_{j9}^{30}(\text{tod}_{i}, \text{dow}^+_{ij}) + f_{j10}^{20}(\text{tod}_{i}, \text{shol}_{ij}) \nonumber \\ & + f_{j11}^{5, 5}(\text{n2ex}_i, \text{tod}_{i}) + f_{j12}^{5, 5}(\text{temp}_{ij}, \text{tod}_{i}) + f_{j13}^{5, 5}(\text{rain}_{ij}^{1/2}, \text{tod}_{i}) + f_{j14}^{10, 10}(\text{doy}_i, \text{tod}_{i})\,\,,
\end{align*}
for $j = 1, \dots, 14$.
Note that, given that the mean model is fitted in a preliminary step to save computational time (see Section \ref{sec:GSP14_eval}), the mean vector fit is exactly the same for each of the models. 

For the \verb|Full| model, each element of the MCD parametrisation of the covariance matrix can be modelled via any of the effects appearing in formula (\ref{mcd_formula}), which we repeat here
\begin{align*}
\eta_{ij} &= \bar{\eta}_{ij} + g_{j1}(t_i) + g_{j2}(t_i^2) + g_{j3}(\text{dow}_{i}) + f_{j1}^{10}(\text{doy}_i) + f_{j2}^{10}(\text{tod}_i) + \nonumber \\  & + \text{wcap}_i \times f_{j3}^{5}(\text{wsp}_{il_j}^{100}) + f_{j4}^{5}(\text{irr}_{il_j}) + f_{j5}^{5}(\text{temp}_{il_j}) + f_{j6}^{5}(\text{rain}_{il_j}^{1/2}) + f_{j7}^{5}(\text{n2ex}_{i})\,\,,
\end{align*}
for $j = 15, \dots, 119$, and where we indicate with $l_j$ the row of $\bold D_i$ or $\bold T_i$ on which the $j$-th linear predictor appears.
Model selection Algorithm \ref{algo:grad_boosting} (see SM \ref{AppB1_PLS}) evaluates, at each step, the gain obtained by adding one of the effects appearing in the formula above to one of the non-zero elements of the $\bold T$ and $\bold D$ factors forming the MCD factorisation. Under the \verb|Cal+Ren| model, the search is restricted to the effects appearing in the formula
\begin{align*}
\eta_{ij} &= \bar{\eta}_{ij} + g_{j1}(t_i) + g_{j2}(t_i^2) + g_{j3}(\text{dow}_{i}) + f_{j1}^{10}(\text{doy}_i) + f_{j2}^{10}(\text{tod}_i) + \nonumber \\  & + \text{wcap}_i \times f_{j3}^{5}(\text{wsp}_{il_j}^{100}) + f_{j4}^{5}(\text{irr}_{il_j})\,\,,
\end{align*}
for $j = 15, \dots, 119$ while, for the \verb|Cal| model, the search is further restricted to the effects in
\begin{align*}
\eta_{ij} &= \bar{\eta}_{ij} + g_{j1}(t_i) + g_{j2}(t_i^2) + g_{j3}(\text{dow}_{i}) + f_{j1}^{10}(\text{doy}_i) + f_{j2}^{10}(\text{tod}_i) \nonumber\,\,,
\end{align*}
for $j = 15, \dots, 119$.
Hence, in the \verb|Full|, \verb|Cal+Ren| and \verb|Cal| models, all the MCD elements are considered but the number of candidate effects is increasingly restricted. 

Under the \verb|Cal+Ren Diag| model, the pool of candidate effects is the same as for \verb|Cal+Ren|, that is those appearing in the formula
\begin{align*}
\eta_{ij} &= \bar{\eta}_{ij} + g_{j1}(t_i) + g_{j2}(t_i^2) + g_{j3}(\text{dow}_{i}) + f_{j1}^{10}(\text{doy}_i) + f_{j2}^{10}(\text{tod}_i) + \nonumber \\  & + \text{wcap}_i \times f_{j3}^{5}(\text{wsp}_{il_j}^{100}) + f_{j4}^{5}(\text{irr}_{il_j})\,\,,
\end{align*}
but $j$ is restricted to take value in $15, 16, \dots, 28$. That is,  Algorithm \ref{algo:grad_boosting} evaluates the gain obtained by adding an effect to a diagonal element of the $\bold D$ factor, while the elements of the $\bold T$ factor are not allowed to vary with the covariates (i.e., each of the corresponding linear predictors contains only an intercept).

\subsection{Details on the Implementation of the Copula-Based Models} \label{AppB2_copula_models}

The \texttt{gaulss+cop}, \texttt{shash+cop} and \texttt{shash+gpd+cop} GAMLSS models \citep{rigby2005generalized} from Section \ref{sec:GSP14_eval} model the marginal distribution of regional net-demand separately from the correlation structure. Here we give more details on the structure of these models and on how they are fitted to data. 

Consider the \texttt{shash+cop} model. Under this model, the net-demand from $j$-th region follows a conditional sinh-arcsinh distribution \citep{jones2009sinh}, controlled by the four-dimensional parameter vector $\boldsymbol \theta_{ij}$. Each element of $\boldsymbol \theta_{ij}$ could potentially be modelled additively but, for the reasons put forward in Section \ref{sec:GSP14_eval}, we let only the log-scale parameter depend on the covariates, and control the location, asymmetry and tail parameters only via intercepts. The \texttt{gaulss+cop} model uses a two-parameter conditional Gaussian model, where the location parameter is kept constant while the log-scale parameter is modelled additively, as explained in Section \ref{sec:GSP14_eval}. 

The predictions of the \texttt{shash+gpd+cop} model are based on the same conditional sinh-arcsinh distribution of \texttt{shash+cop} between quantiles 0.05 and 0.95, and on two separate models based on the generalised Pareto distribution (GPD) beyond them. In particular, let $F_j(\cdot|\bsx_{i})$ be the conditional c.d.f. of the sinh-arcsinh model corresponding to the $i$-th observation and the $j$-th GSP group. Then $q_{ij}^{95} = F^{-1}_j(0.95|\bsx_{i})$ is an estimate of the 95th conditional net-demand percentile under this model. A two-parameter (scale and location) conditional GPD GAMLSS model is fitted to the observed net-demand values that fall above $q_{ij}^{95}$ and a separate GPD model is fitted to those falling below $q_{ij}^{5}$. Given that each of these models is fitted to only around 5\% of the data, we model only the log-scale parameter using a smooth effect of the time of day, while the shape parameter is controlled only by an intercept. Having fitted the sinh-arcsinh and the two GPD models, we build a composite model where the sinh-arcsinh model is used to produce predictions between quantiles 0.05 and 0.95, while the GPD models are used to produce tail predictions.

When fitting the GPD models we impose (via a simple reparametrisation) the constraint that the shape parameter of the distribution, $\xi$, must be larger than 0.001. The reason is that, without this constraint, the estimated shape parameter would in some cases go negative, with disastrous consequences on the test-set performance of the GPD. To see this, recall that on the upper tail we use the GPD to model the distribution of $\epsilon_{hj}^{95} = y_{hj} - {q}^{95}_{hj}$, where $h \in \mathcal{S}_j^{95}$ and $\mathcal{S}_j^{95}$ is a subset of $\{1, \dots, n\}$ such that $y_{hj} \geq {q}^{95}_{hj}$. When the shape parameter $\xi$ of the GPD goes negative, the support of the GPD distribution for $\epsilon_{hj}^{95}$ becomes bounded to $[0, -\sigma/\xi]$, where $\sigma$ is the scale parameter. On the upper tail, some very high out-of-sample net-demand values lead to values of $\epsilon$ that fall outside the support of the distribution and are deemed impossible under this model (the lower tail is affected by the same problem when very low net-demand values occur). Hence the corresponding out-of-sample log-likelihood is undefined, which prevents the computation of the first two scoring rules in Table \ref{tab:GSP14_scores_new_GSP}. More importantly, having a model under which some observed net-demand values are impossible is undesirable. Hence our decision to limit the parameter range. One might worry that the goodness of fit of the GPD-based model might have been compromised by forcing $\xi$ to stay positive. But the goodness of fit diagnostics reported in SM \ref{AppB1_add_results} show that, even with this constraint, the marginal fit of the \texttt{shash+gpd+cop} model is excellent on the in-sample data.

We fit the 14 conditional Gaussian, sinh-arcsinh and (pairs of) GPD models separately to the net-demand of each region, using the fitting methods of \cite{wood2016} and the rolling forecasting approach described in Section \ref{sec:GSP14_eval}. In a second step, the correlation of net-demand across the GSP groups is modelled as follows. Let $F_j(\cdot|\bsx_{i})$ be the conditional c.d.f. of the Gaussian, sinh-arcsinh or of the composition of the sinh-arcsinh and the GPD models corresponding to the $i$-th observation and the $j$-th GSP group. If a marginal net-demand model is well specified, $u_{ij} = F_j(y_{ij}|\bsx_{i})$ should approximately follow a uniform distribution $U(0, 1)$ and $z_{ij} = \Phi^{-1}(u_{ij})$, where $\Phi^{-1}(\cdot)$ is the inverse standard Gaussian c.d.f., should follow a standard Gaussian distribution, $N(0, 1)$. We then adopt a Gaussian copula model by assuming that the $z_{ij}$'s follow a joint multivariate Gaussian distribution, that is
$$
{\boldsymbol z}_i = \begin{pmatrix} z_{i1}  \\
                               z_{i2}\\

                               z_{i3}\\

                               \vdots\\

                             z_{i14}
                             \end{pmatrix} \sim \text{N} \left( \begin{pmatrix} 0   \\
                               0 \\

                               0\\

                               \vdots   \\

                             0 
                             \end{pmatrix}, \begin{pmatrix} 1 & \rho_{1,2} & \rho_{1,3} & \cdots& \rho_{1,14} \\
                               \rho_{1,2} & 1 &  \rho_{2,3} & \cdots& \rho_{2,14}  \\

                               \rho_{1,3} & \rho_{2, 3} & 1& \cdots   & \rho_{3,14} \\

                               \vdots & \vdots & \vdots & \ddots   &  \vdots&  \\

                             \rho_{1,14} & \rho_{2,14} & \cdots &  \rho_{13,14} & 1\\

\end{pmatrix} \right) .
$$

The parameters, $\rho_{1,2}, \dots, \rho_{13,14}$, of the copula model are estimated simply by computing the empirical correlation matrix of the ${\boldsymbol z}_i$ vectors on the training data, using the same rolling forecasting origin used for the multivariate Gaussian models based on the MCD parametrisation, which ensures that all the models are fitted exactly to the same training data. 

The composite models with GAMLSS margins and a static Gaussian copula described above are closely related to the model class discussed in \cite{kock2023truly}.  In particular, \cite{kock2023truly} consider models with GAMLSS margins coupled with a static Gaussian copula for the correlation structure, but fit all parameters jointly via Markov chain Monte Carlo (MCMC) methods, rather than in two steps as done here.

\subsection{Additional Results on UK Regional Net-Demand Forecasting} \label{AppB1_add_results}

Figure \ref{boot_scores_all} shows the bootstrapped differences between the performance scores of several pairs of models, under the scores shown in columns two to seven of Table \ref{tab:GSP14_scores_new_GSP}. The interpretation of the plots is analogous to that of Figure \ref{fig:skills_plot}a.

\begin{figure}
\centering
\includegraphics[scale=0.11]{diff_rest.pdf}
\caption{Bootstrapped differences between the scores of several pairs of models, for the performance score shown in columns two to seven of Table \ref{tab:GSP14_scores_new_GSP}. Negative values mean that the first method is better than the second (e.g., \texttt{Cal+Ren} does better than \texttt{Cal} under all the scores). \textbf{Note}: The numbering of the figures shown here follows from the main text.}
\label{boot_scores_all}
\end{figure}

As mentioned in Section \ref{sec:GSP14_eval_group}, it is possible to linearly transform the joint regional forecasts to obtain marginal probabilistic forecasts of differences in net-demand between regions or macro-regions. Here, we consider the difference in net-demand between the Scottish or South macro-regions and the rest of the country, or between London and its neighbouring regions. These boundaries are of particular interest to the network operator due to the strong influence of wind/solar generation on power flows and to the constraints related to network capacity and stability. Table \ref{tab:scores_diff} reports the performance of each model, when forecasting the marginal distribution of each net-demand difference. As for Table \ref{tab:GSP14_scores_new_GSP}, the log scores of the models based on the sinh-arcsinh distribution are not reported because they are not readily computable.

\begin{table}
\centering
\begin{tabular}{rrrrrrrrr}   &\multicolumn{4}{c}{Scotland - Rest} & \multicolumn{4}{c}{South - Rest} \\
\hline  & Log & CRPS & Pin 001 & Pin 999 & Log & CRPS & Pin 001 & Pin 999 \\    \hline \texttt{Cal} & 2614 & 1305 & 20.75 & 9.31 & 2192 & 945.1 & 18.65 & 5.83 \\    \texttt{Cal+Ren} & 2593 & 1301 & \underline{18.97} & 8.89 & 2168 & 944.1 & 12.76 & 5.80 \\    \texttt{Full} & \underline{2591} & \underline{1299} & 19.32 & \underline{8.36} & \underline{2161} & 943.8 & \underline{12.09} & \underline{5.57} \\    \texttt{Cal+Ren Diag} & 2610 & 1304 & 21.01 & 9.52 & 2183 & 946.3 & 13.46 & 6.67 \\    \texttt{gaulss+cop} & 2597 & 1304 & 19.49 & 9.65 & 2169 & 944.2 & 14.57 & 6.15 \\    \texttt{shash+cop} &  & 1305 & 19.15 & 9.35 &  & 941.8 & 15.08 & 6.34 \\    \texttt{shash+gpd+cop} &  & 1305 & 19.57 & 9.74 &  & \underline{941.7} & 16.46 & 6.09 \\     \hline \end{tabular} 

\vspace{1.25cm}

\begin{tabular}{rrrrrrrrr}   &\multicolumn{4}{c}{London - Neighbours} & \multicolumn{4}{c}{} \\
 \hline  & Log & CRPS & Pin 001 & Pin 999 &   &   &   &   \\    \hline \texttt{Cal} & 932.4 & 367.8 & 8.73 & 2.13 &  &  &  &  \\    \texttt{Cal+Ren} & 899.5 & 365.9 & 7.54 & 2.00 &  &  &  &  \\    \texttt{Full} & 906.3 & \underline{365.7} & 7.84 & \underline{1.93} &  &  &  &  \\    \texttt{Cal+Ren Diag} & 913.6 & 367.1 & 7.64 & 2.23 &  &  &  &  \\    \texttt{gaulss+cop} & \underline{896.8} & 366.7 & 6.96 & 2.45 &  &  &  &  \\    \texttt{shash+cop} &  & 366.7 & \underline{6.62} & 2.44 &  &  &  &  \\    \texttt{shash+gpd+cop} &  & 366.7 & 6.81 & 2.32 &  &  &  &  \\     \hline \end{tabular} 
\caption{Performance scores on 2018 test data, when forecasting the marginal distribution of differences in net-demand between macro-regions. The best score in each column is {\underline{underlined}}. \textbf{Note}: The numbering of the tables shown here follows from the main text.
}
\label{tab:scores_diff}
\end{table}

Due to the importance of the Beast from the East cold wave on the performance scores (see Figure \ref{fig:skills_plot}b), in Tables \ref{tab:GSP14_scores_NO_BEAST}, \ref{tab:GSP14_scores_NO_BEAST_group},  and \ref{tab:scores_diff_NO_BEAST} we report the same scores as in Tables \ref{tab:GSP14_scores_new_GSP}, \ref{tab:GSP14_scores_new_GSP_region}, and \ref{tab:scores_diff}, but after having excluded the Beast from the East period (which is included between the two black marks at the bottom of Figure \ref{fig:skills_plot}b) when fitting the models and evaluating their performance. Looking at Table \ref{tab:GSP14_scores_NO_BEAST}, note that now the multivariate Gaussian \texttt{Cal+Ren} and \texttt{Full} models are better than the non-Gaussian models on the marginal log score as well.

\begin{table}
\centering
\begin{tabular}{rrrrrrrr}   \hline Model & Log & Log Ind & CRPS & Pin 001 & Pin 999 & Var 0.5 & Var 1.0 \\    \hline \texttt{Cal} & -4170 & -1421.3 & 2415 & 20.69 & 22.82 & 8978 & 9421 \\    \texttt{Cal+Ren} & -4440 & -1646.6 & 2408 & 19.61 & 19.56 & 8802 & 9203 \\    \texttt{Full} & \underline{-4444} & \underline{-1694.0} & \underline{2407} & \underline{18.53} & 19.36 & \underline{8770} & \underline{9159} \\    \texttt{Cal+Ren Diag} & -4258 & -1535.9 & 2410 & 20.08 & 20.79 & 8804 & 9207 \\    \texttt{gaulss+cop} & -4161 & -1541.2 & 2412 & 20.20 & 19.46 & 8860 & 9277 \\    \texttt{shash+cop} & -4205 & -1589.8 & 2411 & 20.00 & 18.52 & 8847 & 9269 \\    \texttt{shash+gpd+cop} & -4275 & -1606.6 & 2412 & 19.74 & \underline{18.51} & 8848 & 9279 \\    \hline \end{tabular}
\caption{Performance scores on 2018 test data, when forecasting the joint distribution of net-demand across the 14 GSP groups, after removing the Beast from the East cold wave from the data. The best score in each column is {\underline{underlined}}.}
\label{tab:GSP14_scores_NO_BEAST}
\end{table}

\begin{table}
\centering
\begin{tabular}{rrrrrrrr}  \hline Model & Log & Log Ind & CRPS & Pin 001 & Pin 999 & Var 0.5 & Var 1.0 \\    \hline \texttt{Cal} & 2807 & 3799.2 & 1849 & 14.13 & 18.95 & 2076 & 4429 \\    \texttt{Cal+Ren} & 2724 & 3714.2 & 1844 & 13.09 & 17.72 & 2050 & 4363 \\    \texttt{Full} & \underline{2684} & \underline{3689.3} & \underline{1843} & \underline{12.35} & \underline{17.15} & \underline{2040} & \underline{4342} \\    \texttt{Cal+Ren Diag} & 2818 & 3796.3 & 1847 & 13.63 & 19.89 & 2059 & 4386 \\    \texttt{gaulss+cop} & 2810 & 3752.8 & 1847 & 14.15 & 19.52 & 2053 & 4375 \\    \texttt{shash+cop} &  &  & 1847 & 13.64 & 18.58 & 2052 & 4375 \\    \texttt{shash+gpd+cop} &  &  & 1847 & 13.32 & 19.51 & 2053 & 4379 \\     \hline \end{tabular}
\caption{Performance scores on 2018 test data, when forecasting the joint distribution of net-demand across the five GSP macro-regions, after removing the Beast from the East cold wave from the data. The best score in each column is {\underline{underlined}}.}
\label{tab:GSP14_scores_NO_BEAST_group}
\end{table}

\begin{table}
\centering
\begin{tabular}{rrrrrrrrr}   &\multicolumn{4}{c}{Scotland - Rest} & \multicolumn{4}{c}{South - Rest} \\
\hline  & Log & CRPS & Pin 001 & Pin 999 & Log & CRPS & Pin 001 & Pin 999 \\    \hline \texttt{Cal} & 2439 & 1198 & 11.47 & 9.16 & 2034 & 872.2 & \underline{5.00} & 5.62 \\   \texttt{Cal+Ren} & 2431 & 1196 & 11.46 & 9.02 & 2038 & 872.9 & 5.38 & 5.55 \\    \texttt{Full} & \underline{2427} & \underline{1194} & \underline{11.24} & \underline{8.49} & \underline{2033} & 872.3 & 5.23 & \underline{5.34} \\    \texttt{Cal+Ren Diag} & 2449 & 1199 & 13.62 & 9.45 & 2053 & 875.5 & 6.06 & 6.44 \\    \texttt{gaulss+cop} & 2442 & 1199 & 12.79 & 9.89 & 2035 & 873.0 & 5.46 & 5.89 \\    \texttt{shash+cop} &  & 1200 & 12.78 & 9.59 &  & \underline{871.0} & 5.28 & 5.72 \\    \texttt{shash+gpd+cop} &  & 1200 & 13.22 & 9.84 &  & 871.1 & 5.40 & 5.95 \\     \hline
\end{tabular} 

\vspace{1.25cm}

\begin{tabular}{rrrrrrrrr}   &\multicolumn{4}{c}{London - Neighbours} & \multicolumn{4}{c}{} \\
\hline  & Log & CRPS & Pin 001 & Pin 999 &   &   &   &   \\    \hline \texttt{Cal} & 792.3 & 335.3 & 5.40 & 1.92 &  &  &  &  \\    \texttt{Cal+Ren} & \underline{769.8} & 333.9 & 4.92 & 1.89 &  &  &  &  \\    \texttt{Full} & 770.2 & \underline{333.5} & 4.92 & \underline{1.84} &  &  &  &  \\    \texttt{Cal+Ren Diag} & 784.1 & 334.8 & 5.10 & 2.07 &  &  &  &  \\    \texttt{gaulss+cop} & 777.5 & 335.0 & 5.05 & 2.38 &  &  &  &  \\    \texttt{shash+cop} &  & 334.7 & 4.98 & 2.33 &  &  &  &  \\    \texttt{shash+gpd+cop} &  & 334.9 & \underline{4.75} & 2.23 &  &  &  &  \\     \hline \end{tabular}
\caption{Performance scores on 2018 test data, when forecasting the marginal distribution of differences in net-demand between macro-regions, after removing the Beast from the East cold wave from the data. The best score in each column is {\underline{underlined}}.
}
\label{tab:scores_diff_NO_BEAST}
\end{table}

Table \ref{tab:shape_beast_0} reports the shape parameters of the GPD models for the lower and upper tail, estimated on the whole data as part of the \texttt{shash+gpd+cop} model, with and without the observations corresponding to the Beast from the East cold wave. Note that for many GSP groups this parameter is equal to 0.001. This is because we impose a lower bound of 0.001 on this parameter, as explained in SM \ref{AppB2_copula_models}. 

\begin{table}[htbp]
\centering
\begin{tabular}{rrr|rr}
  \hline
& \multicolumn{2}{c|}{With the Beast from} & \multicolumn{2}{c}{Without the Beast from} \\ 
& \multicolumn{2}{c|}{the East cold wave } & \multicolumn{2}{c}{the East cold wave } \\ 
\hline
 & Lower tail & Upper tail & Lower tail & Upper tail \\ 
  \hline
N Scotland (P) & 0.127 & 0.142 & 0.132 & 0.068\\ 
  S Scotland (N) & 0.133 & 0.145 & 0.132 & 0.135 \\ 
  NE England (F) & 0.001 & 0.001 & 0.001 & 0.001 \\ 
  Yorkshire (M) & 0.007 & 0.001 &  0.018 & 0.001\\ 
  NW England (G) & 0.001 & 0.001 & 0.001 & 0.001\\ 
  Merseyside \& N Wales (D) & 0.001 & 0.114 & 0.001 & 0.015 \\ 
  S Wales (K) & 0.001 & 0.001 & 0.001 & 0.001 \\ 
  W Midlands (E) & 0.044 & 0.092 &  0.043 & 0.030\\ 
  E Midlands (B) & 0.001 & 0.084 & 0.001 & 0.001  \\ 
  E England (A) & 0.001 & 0.039 & 0.001 & 0.001  \\ 
  London (C) & 0.064 & 0.205 & 0.063 & 0.001  \\ 
  SE England (J) & 0.001 & 0.139 & 0.001 & 0.083\\ 
  S England (H) & 0.001 & 0.103 & 0.001 & 0.001 \\ 
  SW England (L) & 0.001 & 0.177 & 0.001 & 0.013 \\ 
   \hline
\end{tabular}
\caption{Estimated shape parameters of the GPD models for the lower and upper tail of the net-demand distribution, for each GSP group.} 
\label{tab:shape_beast_0}
\end{table}

\begin{figure}
\centering
\includegraphics[scale=0.5]{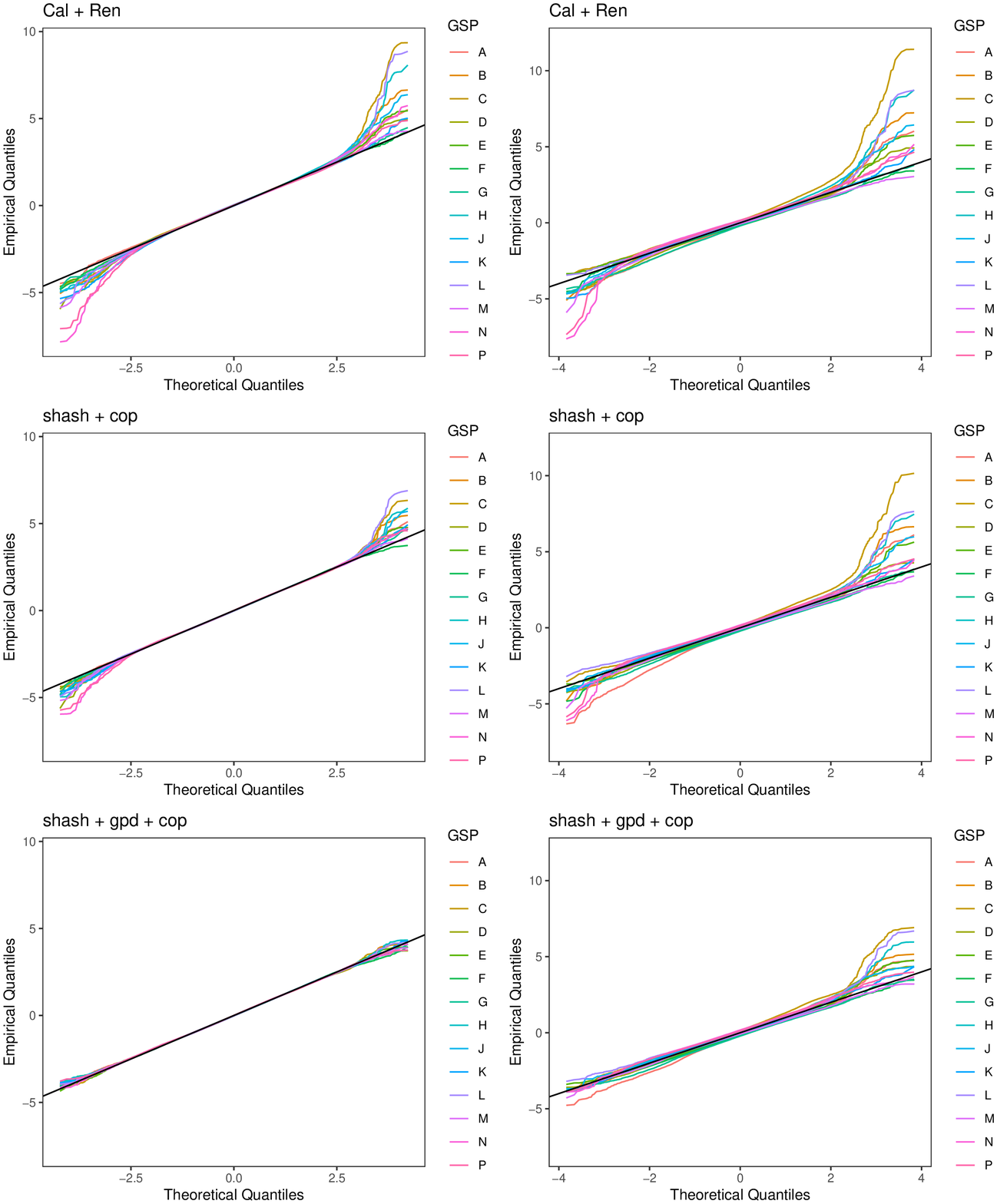}
\caption{Left: QQ-plots of the in-sample quantile residuals for each GSP group, for the Cal+Ren, shash+cop and shash+gpd+cop model. Right: QQ-plots of the out-of-sample (one day ahead) quantile residuals under the same three models.}
\label{fig:qqplots_0}
\end{figure}

\begin{figure}
\centering
\includegraphics[scale=0.5]{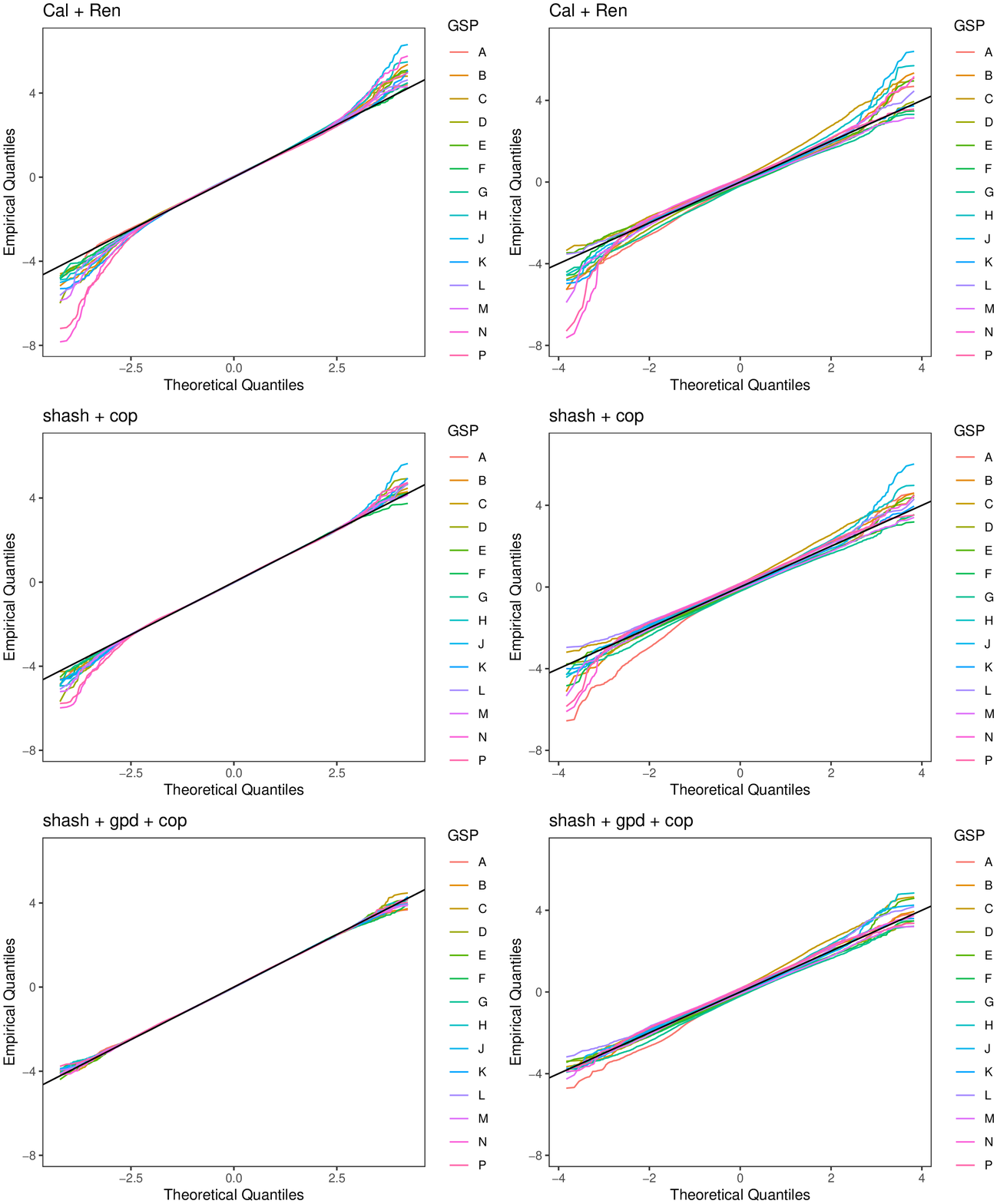}
\caption{Left: QQ-plots of the in-sample quantile residuals for each GSP group, for the Cal+Ren, shash+cop and shash+gpd+cop model. Right: QQ-plots of the out-of-sample (one day ahead) quantile residuals under the same three models. Here the Beast from the East cold wave has been removed from the data. }
\label{fig:qqplot_no_beast_0}
\end{figure}

The QQ-plots shown in Figure \ref{fig:qqplots_0} and \ref{fig:qqplot_no_beast_0} are useful to assess the marginal goodness of fit of three of the models considered in this work. They are based on quantile residuals which, following \cite{dunn1996randomized}, have been computed by evaluating the marginal c.d.f. of each model to obtain uniform residuals and then transforming them via the inverse standard normal c.d.f.. The left column in each figure shows the QQ-plots of the in-sample quantile residuals on 2018 data, obtained after fitting the models to the whole data (2014-18). There is a clear progression, that is the marginal fit gets better and better as the model for the margins becomes more flexible. The right column in each plot shows QQ-plots of the out-of-sample (one day ahead) quantile residuals on 2018 data, obtained under the monthly model updating scheme described in Section \ref{sec:GSP14_eval}. We can still see a progression as we move to more flexible marginal models, but this is much less clear than when looking at in-sample residuals.  Hence, the goodness of fit improvements brought about by adopting more flexible marginal models are more limited in the test set, which is not surprising considering the extraordinary nature of events such as the Beast from the East cold wave.

\newpage 

\newpage 
\putbib
\end{bibunit}

\end{document}